\documentclass[useAMS,usenatbib]{mn2e}
\usepackage{graphicx}
\usepackage{wrapfig}
\makeatletter

\newcommand{\Rmnum}[1]{\expandafter\@slowromancap\romannumeral #1@}
\makeatother

\title[Galaxy environments]{Galaxy And Mass Assembly (GAMA): Galaxy environments and star formation rate variations}
\author[D. B.Wijesinghe et al.]
{D. B. Wijesinghe$^{1}$\thanks{E-mail:D.Wijesinghe@physics.usyd.edu.au},
A. M. Hopkins$^{2}$, S. Brough$^{2}$, E. N. Taylor$^{1}$, P. Norberg$^{3}$, 
\newauthor
 A. Bauer$^{2}$, M. J. I. Brown$^{4}$, E. Cameron$^{5}$, C. J. Conselice$^{6}$, S. Croom$^{1}$, S. Driver$^{7,8}$,
\newauthor
M. W. Grootes$^{9}$, D. H. Jones$^{4}$, L. Kelvin$^{7,8}$, J. Loveday$^{10}$, K. A. Pimbblet$^{4}$, 
\newauthor
C. C. Popescu$^{11}$, M. Prescott$^{12}$, R. Sharp$^{13}$, I. Baldry$^{12}$, E. M. Sadler$^{1}$, J. Liske$^{14}$, 
\newauthor
A. S. G. Robotham$^{7,8}$, S. Bamford$^{6}$, J. Bland-Hawthorn${^1}$, M. Gunawardhana$^{1}$, 
\newauthor
M. Meyer$^{7}$, H. Parkinson$^{3}$, M. J. Drinkwater$^{15}$, J. Peacock$^{3}$, R. Tuffs$^{9}$,\\ 
$^{1}$Sydney Institute for Astronomy, School of Physics, University of Sydney, NSW 2006, Australia\\
$^{2}$Australian Astronomical Observatory, PO Box 296, Epping, NSW 1710, Australia\\
$^{3}$Institute for Astronomy, University of Edinburgh, Royal Observatory, Blackford Hill, Edinburgh EH9 3HJ, UK\\
$^{4}$School of Physics, Monash University, Clayton, Victoria 3800, Australia\\
$^{5}$ETH Zurich, Institute for Astronomy, HIT J12.3, CH-8093 Zurich, Switzerland\\
$^{6}$School of Physics \& Astronomy, University of Nottingham, University Park, Nottingham NG7 2RD, UK\\
$^{7}$International Centre for Radio Astronomy Research (ICRAR), University of Western Australia, Crawley, WA 6009, Australia\\
$^{8}$Scottish Universities' Physics Alliance (SUPA), School of Physics and Astronomy, University of St Andrews,\\
North Haugh, St. Andrews, KY16 9SS, UK\\
$^{9}$Max-Planck-Institut f\"ur Kernphysik, Saupfercheckweg 1, 69117, Heidelberg, Germany\\
$^{10}$Astronomy Centre, University of Sussex, Falmer, Brighton BN1 9QH\\
$^{11}$Jeremiah Horrocks Institute for Astrophysics and Supercomputing, University of Central Lancashire, Preston, PR1 2HE, UK\\
$^{12}$Astrophysics Research Institute, Liverpool John Moores University, Twelve Quays House, Egerton Wharf, Birkenhead, CH411LD, UK\\
$^{13}$Research School of Astronomy \& Astrophysics, the Australian National University, Cotter Road, Weston Creek, ACT 2611, Australia\\
$^{14}$European Southern Observatory, Karl-Schwarzschild-Str.~2, 85748, Garching, Germany\\
$^{15}$Department of Physics, University of Queensland, QLD 4072, Australia\\}
\begin{document}

\date{Accepted 2011 November}

\pagerange{\pageref{firstpage}--\pageref{lastpage}} \pubyear{2009}

\maketitle

\label{firstpage}

\begin{abstract}
We present a detailed investigation into the effects of galaxy environment on their star formation rates (SFR) using 
galaxies observed in the Galaxy and Mass Assembly Survey (GAMA). We use three independent volume-limited samples of galaxies
within $z < 0.2$ and $M_{r} < -17.8$.
We investigate the known SFR-density relationship and explore in detail the dependence of SFR on stellar mass and density.
We show that the SFR-density trend is only visible when we include the passive galaxy population along with the star-forming population.
This SFR-density relation is absent when we consider only the star-forming population of galaxies, consistent with previous work.
While there is a strong dependence of the $\mathrm{EW}_{\mathrm{H}\alpha}$ on density we find, as in previous studies, that these trends are largely due to the passive
galaxy population and this relationship is absent when considering a ``star-forming'' sample of galaxies.
We find that stellar mass has the strongest influence on SFR and $\mathrm{EW}_{\mathrm{H}\alpha}$ with the environment having no significant effect on the star-formation properties of the star 
forming population. We also show that the SFR-density relationship is absent for both early and late-type star-forming galaxies. 
We conclude that the stellar mass has the largest impact on the current SFR of a galaxy, and any environmental effect is not detectable. The observation that the trends with density are due 
to the changing morphology fraction with density implies
that the timescales must be very short for any quenching of the SFR in infalling galaxies.
Alternatively galaxies may in fact undergo predominantly in-situ evolution where
the infall and quenching of galaxies from the field into dense environments is not the dominant evolutionary mode. 

 
\end{abstract}

\begin{keywords}
galaxies: evolution -- galaxies: environment -- galaxies: general -- galaxies: formation
\end{keywords}

\section{Introduction}

The factors affecting the evolution of galaxies are unlikely to be limited to their intrinsic properties, as it has been widely reported that star formation rate (SFR)
is suppressed in cluster environments compared to SFRs in field galaxies \citep{Lws:02,Gom:03, vdB:08,Pas:09}.
In contrast, Baldry et al. (2004), Balogh et al. (2004), Peng et al. (2010), Bolzonella et al. (2010) and Li et al. (2011), among others, have shown that the SFR-environment relation is driven 
largely by the changing fractions of passive galaxies and the SFR-mass relation. More recently evidence has emerged to show that not only does the changing fraction of passive galaxies 
indicate SFR suppression but that it can be used to constrain the time-scale of SFR-suppression (Wolf et al. 2009, Vulcani et al. 2010, Li et al. 2011, von der Linden et al. 2010, Macgee et al. 2011).


Evidence for the suppression of star-formation in galaxies at the cores of clusters is widely accepted through the work of many authors \citep[e.g.,][]{Bal:97, Hmt:98, Pog:99, Cch:01}.
\citet{Hmt:98} also found a continuous correlation with local galaxy density where the SFR decreases with increasing density. This is important as only a small 
fraction of galaxies occupy the cores of clusters, and in order to make judgments regarding the broad evolution of galaxies we must understand the effect density has on galaxies in varying 
positions in the clustering environment.

This idea was expanded by \citet{Lws:02} and G\'{o}mez et al. (2003). \citet{Lws:02}, using the 2dF Galaxy Redshift Survey,
found that SFR depends strongly on local density and is independent of proximity to a rich cluster. G\'{o}mez et al. (2003), using the Sloan Digital Sky Survey
(SDSS) Early Data Release, showed that the
SFR-density relation is most visible for galaxies with the highest SFRs.
A similar conclusion was reached by \citet{Pmb:06} using the [O{\sc ii}] line to measure the SFR for clusters at $z \approx 3$.

These results lead to proposals that the stripping of hot gas reservoirs from galaxies during hierarchical formation, where galaxies are accreted into high density regions from low density regions 
\citep{KWG:93,Col:00}, is the reason behind the reduction in SFRs in high density regions. There remains a large uncertainty behind the exact physical
mechanisms that lead to this reduction of the gas in galaxies. Processes such as interactions between intragalactic and intergalactic media \citep{GG:72}, suppression of the accretion of gas 
rich materials \citep{LTC:80,Bek:01,Bek:02,Bek:09}, tidal interactions \citep{BV:90}, galaxy harassment through high velocity encounters with other 
galaxies \citep{Zab:96, Mor:96}, ram pressure stripping of the cold gas \citep{GG:72,QMB:00} and in-fall and quench (also known as strangulation)
have all been suggested as the underlying mechanism in 
SFR suppression. The significance of these processes for SFR reduction in clusters still remains speculative though, requiring more detailed analysis to uncover the fundamental processes that drive 
this SFR suppression.

\citet{Bal:97} compared galaxies with similar luminosities and morphological features such as the bulge-to-disk ratio in cluster and field galaxies and found that SFRs were lower in
cluster galaxies. This evidence suggests that the SFR-density
relation cannot be explained using only the density-morphology relation \citep{Drs:80,vdW:08,Bam:09}. \citet{Lws:02} reached a similar conclusion with findings that showed that the correlation between SFR
and density predicted by the density-morphology relation is weaker than observed.

There has also been evidence for spirals in clusters that show SFRs similar to or even greater than spirals in the field \citep{GJ:85,MW:93,Gav:98}. The first
observations of star-forming galaxies in clusters were made by \citet{BO:78} and \citet{BO:84}. \citet{DG:83} concluded that an epoch of strong star-formation had 
recently ended in these galaxies. \citet{GJ:85} indicated that these galaxies may be undergoing a transient phenomenon before undergoing SFR suppression. 

Interaction with the intracluster medium \citep{GG:72},
galaxy-galaxy interactions \citep{LH:88}, gas compression by ram pressure \citep{DG:83,Vol:01} and tidal interactions \citep{MW:00} have been proposed
as the root causes for a burst in SF prior to quenching it. \citet{MS:04} argue that tidal interactions are a likely candidate for
this behavior as they are likely to cause both the burst and suppression in SFRs. There has also been evidence for the ``pre-processing'' of galaxies well before they enter the high density
regions of clusters. The burst of star-formation observed in late-type galaxies as they are accreted into filaments is one such effect \citep{Pot:08}.

\citet{Elb:07} showed that the SFR-density relation inverts by redshift
$\sim$\,1, consistent with the cluster galaxies having enhanced star-formation in the past compared to field galaxies, and also consistent with the idea that these galaxies form earlier and
form their stars more rapidly at earlier times. Therefore, this effect is partly an evolutionary one, as implied by the Butcher-Oemler effect 
\citep{BO:78,BO:84} where the blue-fraction of galaxies in clusters increases with redshift. \citet{Grt:11a} and \citet{Grt:11b} also show that mass and SFR 
are tightly coupled up to a redshift of $z \approx 1$, or even higher, with environment having a slight dependence.

Evidence has also emerged in the past few years that the apparent SFR-density relation is a result of the changing fractions of early and late-type galaxies with increasing density.
One of the first to point this out was Baldry et al. (2004) who use galaxy colours to show that the colour-mass and colour-concentration index relation are not strong
functions of environment. Balogh et al. (2004) also identified that the fractions of star-forming and quiescent galaxies vary strongly with density.
Baldry et al. (2006) do, however, identify that the fraction of galaxies in the red sequence have a substantial dependence on stellar mass and environment.

More recently, Bolzonella et al. (2010), Wetzel et al. (2011), Deng et al. (2011) and Lu et al. (2011) use SFR and specific SFR (SSFR) to show that the SFR-density relation is largely a product
of the changing fraction of passive galaxies as well as the relation between SFR and mass. Deng et al. (2011), for instance, show that the environmental dependence of SFRs and SSFRs is much 
stronger for red galaxies compared to blue galaxies implying that an increasing fraction of red galaxies is a primary driver for the SFR and SSFR-density relation.

The question then is whether this evidence for changing fraction of galaxy types indicates that there is no quenching of the SFR of galaxies falling into clusters or the quenching occurs on 
a very rapid time-scale not observed in the above analyses. Peng et al. (2010) argue that the dependence of the red fraction of galaxies on environment for fixed masses is indeed evidence for
the suppression of SFRs with increasing density. They demonstrate that stellar mass and environment affect passive and star-forming galaxies in different and independent ways
which they refer to as ``mass quenching'' and ``environmental quenching''. While the ``mass quenching'' is a continuous process that is proportional to the SFR of the galaxies, the 
``environmental quenching'' occurs on a very short time-frame that could possibly be a result of satellite galaxies falling into larger halos.

There is controversy regarding the timescales on which the ``infall and quench'' model operates.
Wolf et al. (2009) and Vulcani et al. (2010) present evidence for SFR suppression in star-forming galaxies in clusters, particularly for low-mass galaxies, reasoning that the quenching
process occurs over longer timescales due to the abundance of red spiral galaxies and the unchanged SFR-mass relation. Similarly, \citet{vdL:10} suggested that the quenching timescales are longer
and comparable to the cluster-crossing time on the scale of a few Gyr.
Balogh et al. (2004) and McGee et al. (2011), however, find no evidence for changes in colour or SFR with density for 
star-forming galaxies. McGee et al. (2011), therefore, argues that the processes that lead to the SFR suppression must be fast-acting, and propose infall and quench as a possible quenching 
mechanism. Bolzonella et al. (2010), Wetzel et al. (2011) and Weinmann et al. (2010) agree that the lack of SFR-density relation in star-forming 
galaxies means that these time-scales are likely to be short.

In our analysis, we confirm the result from recent work showing that the previously reported ``suppression'' in SFR with increasing local density can be explained
as a consequence primarily of the changing population mix of galaxies.
We further demonstrate that the range of SFRs for actively star-forming galaxies is independent of local density. To add clarity in the discussion of these
effects, we will limit our use of the term ``suppression'' only to cases when we refer to physical mechanisms that may cause the SFR to be reduced. Otherwise we restrict ourselves to describing the
distribution of SFRs independently of any implied physical mechanism promoting or retarding them.

In $\S$\,2 we describe the data used in this analysis. In $\S$\,3 we compare the SFR-density relation for a sample of the GAMA galaxies with the results of \citet{Gom:03}.
The interplay between stellar mass, SFR, $\mathrm{EW}_{\mathrm{H}\alpha}$, and density is explored in detail in $\S$\,4, and in $\S$\,5 we illustrate the lack of density dependence in the
distribution of SFR for a sample of star-forming galaxies. We quantify galaxy morphologies into early and late-types in $\S$\,6, to tease apart further this lack of density
dependence for SFR, before discussing our results in $\S$\,7 and concluding in $\S$\,8.
Throughout we assume $H_{0}=$70\,km\,s$^{-1}$\,Mpc$^{-1}$, $\Omega_{M}=0.3$ and $\Omega_{\Lambda}=0.7$.
All magnitudes are in the AB system.

\section{Data}
We use data from the Galaxy and Mass Assembly (GAMA) survey
\citep{Drv:09, Drv:11}. GAMA is a multi-band imaging and spectroscopic survey covering $\approx$144 square degrees of sky in three
$12^{\circ} \times 4^{\circ}$ regions \citep{Rob:09, Bld:09, Drv:11}.
The original spectroscopy comes from the AAOmega spectrograph \citep{Shp:06} at the Anglo-Australian Telescope (AAT).

\begin{figure}
\centerline{\includegraphics[width=65mm, height=65mm]{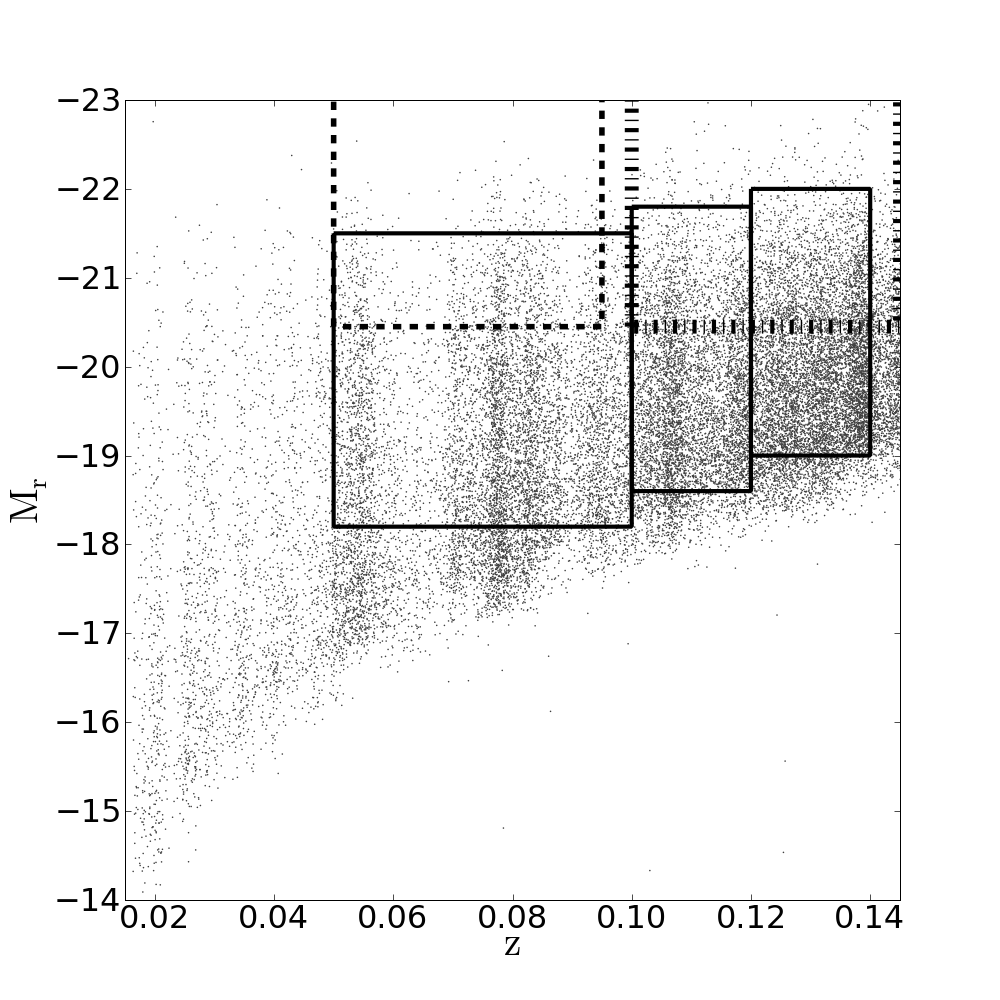}}
\caption{The absolute $r$-band magnitude as a function of redshift. The three volume-limited samples used for our analysis are represented by the solid lines and are defined in Table 1.
The volume-limited region used by G\'{o}mez et al. (2003) is displayed in the region $M_{r} < -20.45$ and $0.05 < z < 0.095$ (dashed line) and our equivalent volume-limited region is 
displayed in the region $M_{r} < -20.45$ and $0.10 < z < 0.145$ (dashed-dot line).}
\label{z_vs_r}
\end{figure}

Three volume-limited samples were selected for the main analysis defined in Table 1 and as shown in Figure~\ref{z_vs_r}. We use multiple volume-limited samples in order to 
assess the effects of the SFR-density relation over a range of possible redshifts and $\mathrm{M}_{r}$ values. The discontinuity in the redshift ranges for the volume-limited samples is to ensure 
that the widest range of redshifts is covered as well as to avoid effects due to sky-lines that fall on the H$\alpha$ wavelength when it is redshifted between z\,=\,0.145 and 0.175
\citep{Gun:11}. 
In order to minimise any evolutionary effects that might otherwise bias the results, each volume-limited sample was chosen to have a reasonably narrow 
redshift range. A fourth volume-limited sample was constructed in order to carry out an analysis comparable 
to that by G\'{o}mez et al. (2003). As our data does not extend to the same $r$-band absolute magnitude ($M_{r}$) values at the redshifts investigated by G\'{o}mez et al. (2003),
due to the smaller survey area of GAMA, we opt 
to use a similar limit in $M_{r}$ at a slightly higher redshift, with $M_{r} < -20.45$ and $0.10 < z < 0.145$. This limit was chosen to have the same width in redshift as \citet{Gom:03}.
In comparison the G\'{o}mez et al. (2003) volume-limited sample was defined as 
$M_{r} < -20.45$ and $0.05 < z < 0.095$. Both these regions are displayed in Figure~\ref{z_vs_r}, but only the volume-limited sample defined as $M_{r} < -20.45$ and $0.10 < z < 0.145$
from GAMA is used in the analysis below.

\begin{table}
\centering
\caption{volume-limited samples}
\begin{tabular}{cccc} 
\hline\hline \\
Label & z & $M_{r}$ & Sample size  \\ [0.5ex]
\hline \\
 VL1 & $0.05 < z < 0.01$ & $-21.5 < M_{r} < -18.2$ & 6036\\
 VL2 & $0.10 < z < 0.12$ & $-21.8 < M_{r} < -18.6$ & 4421\\
 VL3 & $0.12 < z < 0.14 $ & $-22.0 < M_{r} < -19.0$ & 6133\\[1ex]
\hline
\end{tabular}
\label{table:vl}
\end{table}

The SFRs are calculated using the prescription outlined in \citet{Wij:11}. The SFRs are derived using the H$\alpha$ EW with stellar absorption 
corrections of 0.7\AA\ \citep{Gun:11,Bru:11}, and the dust obscuration corrections are calculated using the Balmer decrement.
Stellar masses used in the analysis were derived by \citet{Tal:11} using the models of \citet{BC:03} with exponentially
declining SFHs, a Chabrier IMF \citep{Chb:03}, and Calzetti dust \citep{Cal:01}.

The galaxy environments are defined for a volume-limited sample of galaxies with $M_{r} < -20$ and $z < 0.18$ (Brough et al. in prep.; Rowlands et al. 2011). 
The density, $\Sigma_{N}$, is calculated using a $5^{th}$
nearest neighbour metric which is similar to $\Sigma_{1}$ in \citet{Bru:11} and $\Sigma_{4.5}$ used in \citet{Prs:11}. 
The surface density ($\Sigma_{N}$), in number of galaxies per $\mathrm{Mpc}^{2}$, to the $5^{th}$ nearest neighbour is calculated as,
\begin{equation}
\Sigma_{5} = \frac{N}{\pi d^{2}_5}
\end{equation}
where $d_{5}$ is the projected comoving distance to the $5^{\mathrm{th}}$ nearest neighbour within $\pm 1000 \mathrm{kms}^{-1}$.
Densities are calculated for galaxies with $r_{petro} \leq 19.4$ (where $r_{petro}$ is the $r$-band petrosian magnitude), $0.01 < z < 0.18$ and with reliable redshifts (nQ $\ge 3$; Driver 
et al. 2011). All densities are given in units of number of galaxies per $\mathrm{Mpc}^{2}$. We use bins with a fixed number of galaxies when determining density bins for the Figures below.
The number of galaxies per bin is given explicitly in the caption or description for each Figure as this number varies according to the overall number of galaxies available for the analysis.

Active galactic nuclei (AGN) were eliminated from the analysis using the emission line diagnostic ratio prescription of \citet{Kwl:06}.
We do not expect our results to be sensitive to how we select AGNs \citep{Bam:08}.
We use two sample of galaxies, ``full'' and ``star-forming'', neither of which include AGNs.
The ``full'' sample only excludes galaxies with SFR $> 10^{3} M_{\sun} yr^{-1}$.
We exclude these galaxies due to their extreme SFRs that may result from overestimated obscuration or aperture corrections.
Our results are not changed substantially if this extreme population is retained.
The volume-limited samples, VL1, VL2 and VL3 contain only 3, 5 and 10 galaxies with SFR $> 10^{3} M_{\sun} yr^{-1}$ respectively.

The ``star-forming'' sample is a subset of the ``full'' sample.
We select H$\beta$ equivalent width (EW) $> 1.5$\,\AA\ and $f_{\mathrm{H}\alpha}$ / $f_{\mathrm{H}\beta}$ $< 15$ in order to eliminate low signal-to-noise (S/N) 
H$\beta$ fluxes and also galaxies with H$\alpha$ fluxes that contained discrepancies in their measurements respectively.
These selections ensure that the spectra used to measure the SFRs, EWs and dust obscuration corrections are robust.
Positive EWs represent emission. We also make a further selection cut of SFR $> 10^{-3} M_{\sun} yr^{-1}$ to ensure these galaxies are indeed star-forming.
The three volume-limited samples, VL1, VL2 and VL3 contain 3187, 1851 and 2298 star-forming galaxies respectively which comprise the ``star-forming'' sample for each of these volume-limited samples.

We may still be missing a population of high SFR galaxies with faint broadband emission (fainter than the GAMA magnitude limit, see Gunawardhana et al. in prep.).
This bias, however, is also true of the \citet{Gom:03} sample, which we compare to below. We return to possible implications from this effect for our results in the discussion.

\section{Suppression or Population Mix?}
	
\subsection{Suppression}

The results of \citet{Lws:02} and G\'{o}mez et al. (2003) have shaped our understanding of galaxy evolution and environmental effects over the past decade. 
In order to examine the relationship between density and SFR  we begin by reproducing the analysis of \citet{Gom:03}. 
We aim to investigate a sample as close as possible to being directly comparable to this previous work in order to establish a baseline
for further exploration. We compare the results of G\'{o}mez et al. (2003) to galaxies in a similar $M_{r}$ range, although 
at a higher redshift, in order to have a significant sample size. We compare with the G\'{o}mez et al. (2003) analysis as opposed to that of \citet{Lws:02} simply 
because the density measurement of G\'{o}mez et al. (2003) is more comparable to ours.

The results of G\'{o}mez et al. (2003) were derived for a sample of 8598 galaxies with densities derived using a $10^{\mathrm{th}}$ nearest neighbour approach.
Our comparison volume-limited region contains 5019 galaxies with densities measured using the distance to the $5^{\mathrm{th}}$ nearest neighbour approach
with $M_{r} < -20.45$ and $0.10 < z < 0.15$.
These densities, while not quantitatively comparable, are nonetheless quite similar (they probe the same local environment), and are sufficient for the comparison we present here.

For this comparison sample we derived the 
$\mathrm{EW}_{\mathrm{H}\alpha}$ and the SFR in a manner similar to that of G\'{o}mez et al. (2003) and we use the ``full'' sample of galaxies. 
The $\mathrm{EW}_{\mathrm{H}\alpha}$ values for our comparison sample are simply the observed values and no corrections have been applied. The H$\alpha$ SFRs derived for 
these galaxies only include dust corrections (using the formalism of Hopkins et al. 2001) and do not include aperture corrections, again following
G\'{o}mez et al. (2003). The SFRs, however, were derived using the $\mathrm{EW}_{\mathrm{H}\alpha}$ and the $r$-band absolute magnitude values as opposed to the $f_{\mathrm{H}\alpha}$ as 
in \citet{Gom:03}.

\begin{figure}
\begin{center}
\includegraphics[width=58mm, height=58mm]{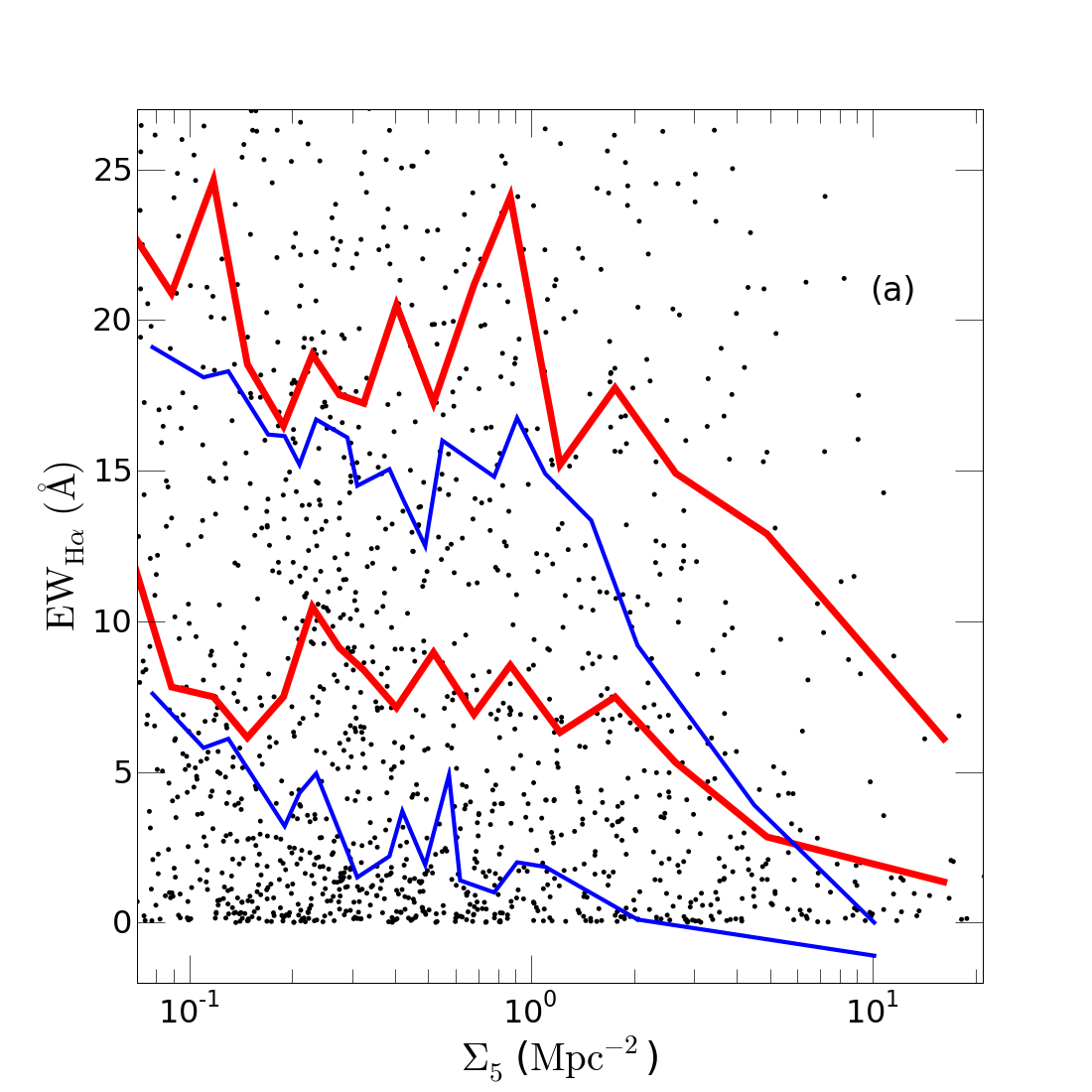}
\includegraphics[width=58mm, height=58mm]{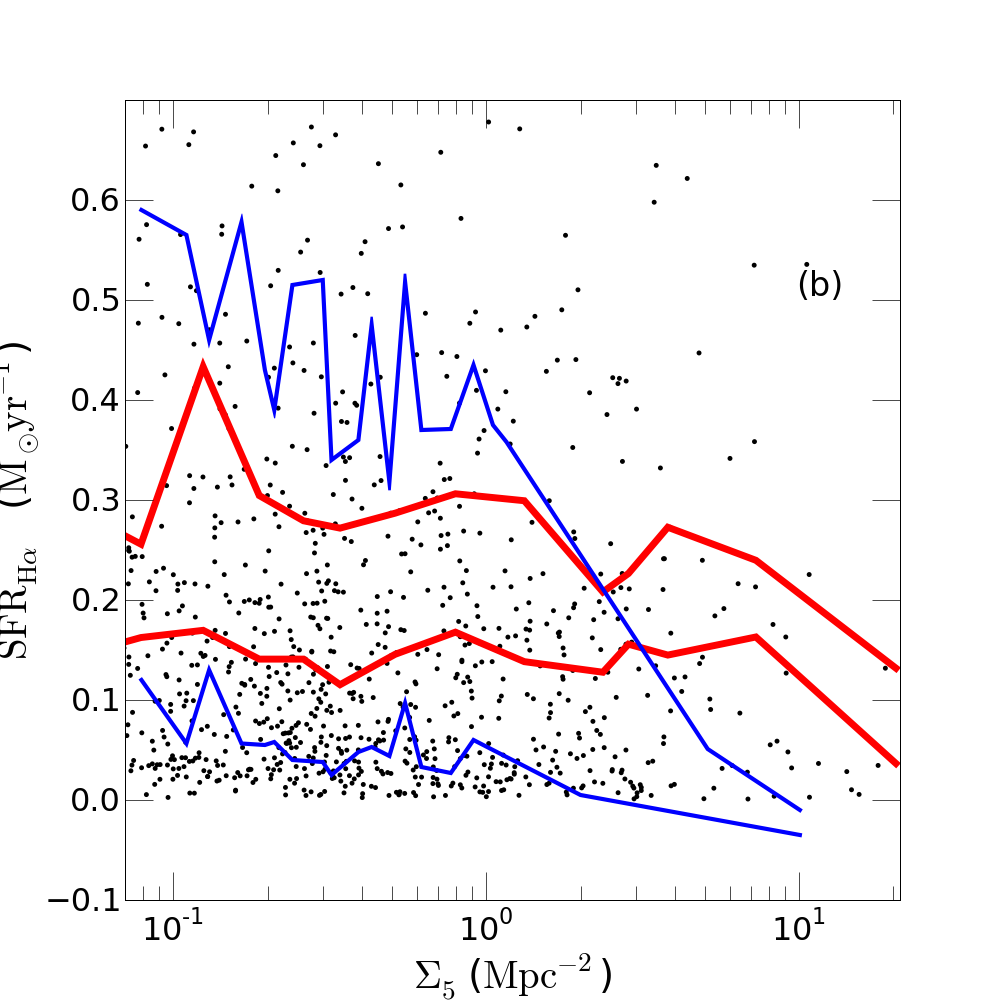}
\caption{(a) $\mathrm{EW}_{\mathrm{H}\alpha}$ as a function of density for the sample of G\'{o}mez et al. (2003) (thin blue line) and for our ``full'' sample (thick red line and black data points) 
showing the 75$^{th}$ and median percentiles. (b) $\mathrm{SFR}_{\mathrm{H}\alpha}$ as a function of density.
There is a significant relationship between  $\mathrm{EW}_{\mathrm{H}\alpha}$ and density for the sample in panel (a), and a high degree of consistency with the result of \citet{Gom:03}.
Panel (b) also shows a trend between $\mathrm{SFR}_{\mathrm{H}\alpha}$ and density, however, this is not as strong as that seen in panel (a).
Panel (a) includes 1953 galaxies from the volume-limited sample and uses bins of 100 galaxies. 
Panel (b) includes 1962 galaxies for which SFR estimates can be made. The trend in panel (b) uses bins of 100 galaxies up to a density of 2 Mpc$^-2$ and uses bins of 25 galaxies for the 
following bins. The reduction in the number of galaxies in the high density bins is due to the limited number of galaxies available at high densities for this sample.
These figures show that both our sample and the sample of \citet{Gom:03} share similar SFR-density and EW-density trends.}
\end{center}
\label{fig:density_ha}
\end{figure}

Figure 2 compares our measurements against those from G\'{o}mez et al. (2003). 
Lines representing the median and $75^{\mathrm{th}}$ percentiles are shown.
Figure 2\,(a) shows a very similar trend between our sample (red line) and that of G\'{o}mez et al. (2003) (blue line). The $\mathrm{EW}_{\mathrm{H}\alpha}$ of our sample, however,
is consistently higher than that of the G\'{o}mez et al. (2003) sample. 
This is a consequence, in part, of the fact that we have measured
EWs only for emission lines. Moreover, galaxies with H$\alpha$ flux values
below $10^{-17} \mathrm{Wm}^{2}$ were excluded from the analysis, as these correspond
to low signal-to-noise lines whose equivalent widths are highly
uncertain.

Figure 2\,(b) compares the SFR distribution. 
The result shows a weaker, but still significant, trend to that of G\'{o}mez et al. (2003), and our SFRs
appear to be lower by about a factor of two. This may be a consequence of our different density estimator, combined with our use of EW rather than flux in the calculation of SFR here. 
Probably the larger effect, though, is that while we sample the brightest $M_{r}$ values in GAMA, the smaller survey area of GAMA means that with the same absolute magnitude
limit as \citet{Gom:03} we are biased towards low SFRs.
As a consequence of not sampling galaxies as bright as those of \citet{Gom:03} we expect to be dominated by lower SFRs, as seen in Figure 2\,(b).


\begin{figure}
\begin{center}
\includegraphics[width=58mm, height=58mm]{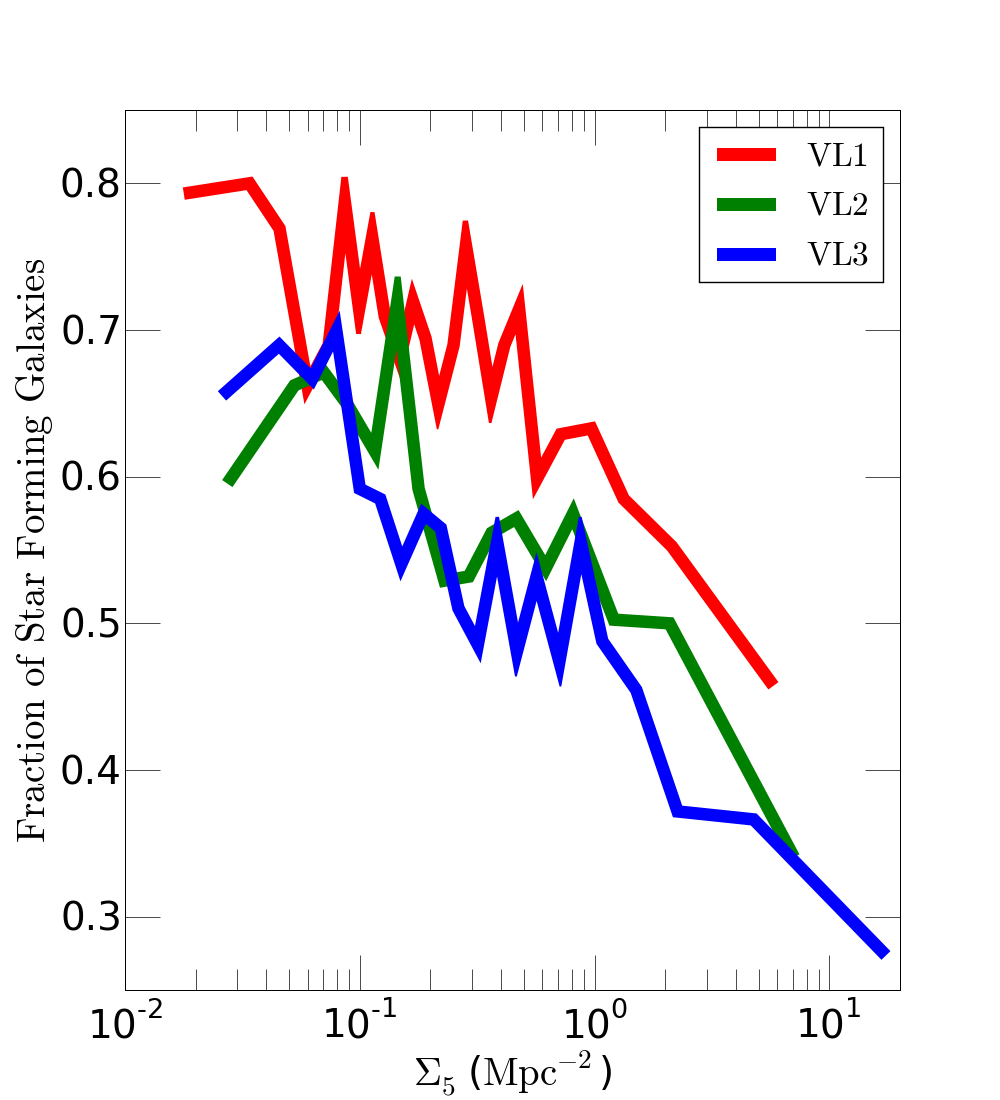}
\caption{Star-forming galaxy fraction as a function of density for the three volume-limited samples, with VL1, VL2 and VL3 represented by the red, green and blue lines respectively.
We use adaptive density bins with the condition that for each density bin the ``star-forming'' sample must contain at least 100 galaxies.
All three samples clearly show a decrease in the fraction of star-forming galaxies at increased densities.}
\end{center}
\label{fig:density_sf_frac}
\end{figure}

\subsection{Population Mix}

While the comparison with the analysis of G\'{o}mez et al. (2003) provides a clear indication that our sample follows the trends previously established, an analysis of the star-forming galaxy 
fraction of the three volume-limited samples is illuminating.
We define the star-forming fraction as the ratio of the number of galaxies in the ``star-forming'' sample to the ``full'' sample.
We use adaptive density bins requiring the ``star-forming'' sample to contain a minimum of 100 galaxies per bin.
Figure 3 shows a systematic decrease in the fraction of star-forming galaxies with increasing density for the three volume-limited samples.
The fraction drops from $\sim$60\% to $\sim$10\% over our observed density range (Figure 3).
The red, green and blue lines correspond to a redshift range of  $0.05 < z < 0.01$, $0.10 < z < 0.12$ \& $0.12 < z < 0.14$ respectively, as defined
in Table 1. 

The density ranges over 3 orders of magnitude for each volume limited sample from $\approx 10^{-2}$ to $\approx 10^{1}$.
There is mild variation with redshift, with the overall star-forming galaxy fraction being lower with increasing redshift. 
This is likely to be related to 
galaxy stellar mass, as we are less sensitive to lower stellar mass galaxies at higher redshift. As high stellar mass galaxies are more likely to be passive, 
we would expect to see a lower star-forming galaxy fraction at higher redshift.

\begin{figure*}
\centerline{\includegraphics[width=58mm, height=58mm]{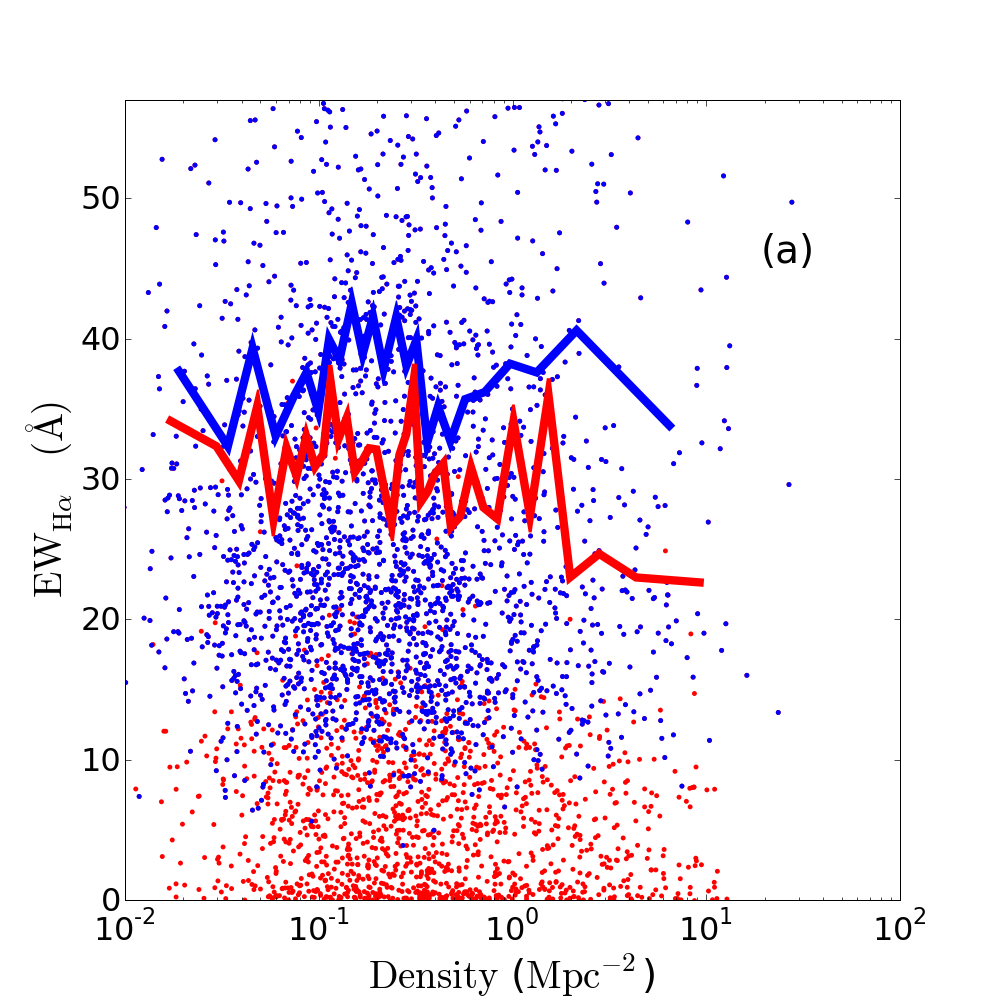}
\includegraphics[width=58mm, height=58mm]{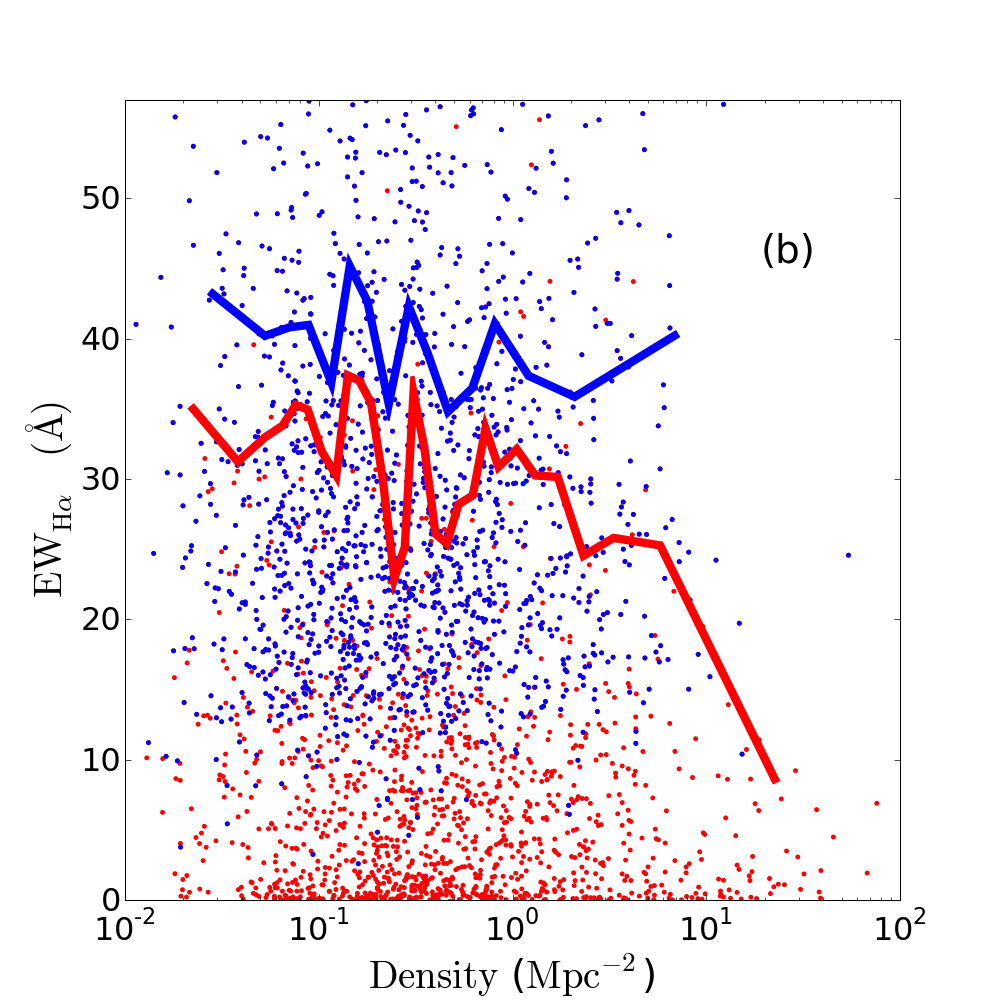}
\includegraphics[width=58mm, height=58mm]{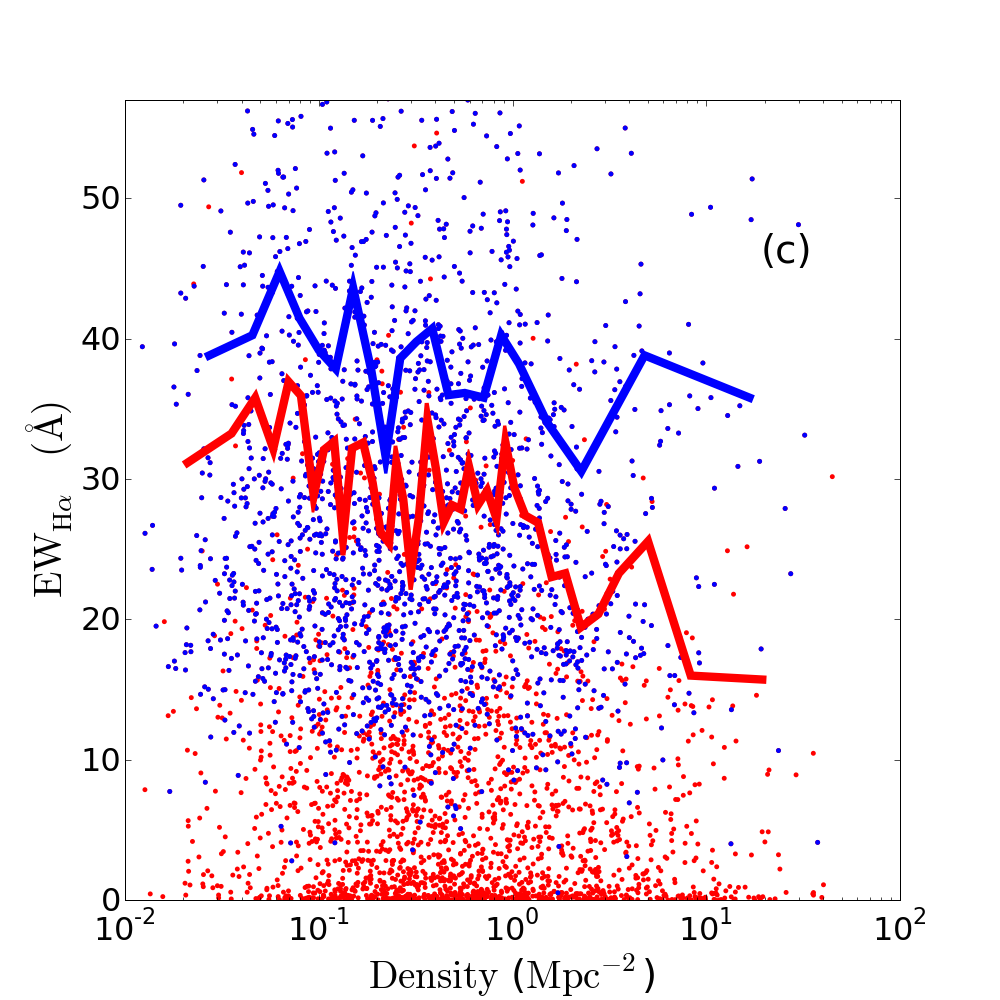}}
\caption{$\mathrm{EW}_{\mathrm{H}\alpha}$ as a function of density for VL1, VL2 and VL3 (panels a, b \& c respectively). In each panel the blue (red) dots and line correspond to the ``star-forming'' 
(``full'') sample of galaxies. The lines show the 75$^{th}$ percentile of the $\mathrm{EW}_{\mathrm{H}\alpha}$ distribution.
This Figure shows that the EW-density relation observed in 
Figure~\ref{fig:density_ha} only occurs in the ``full'' sample of galaxies and not the ``star-forming'' sample.}
\label{fig:density_haew}
\end{figure*}

The relationship between $\mathrm{EW}_{\mathrm{H}\alpha}$ and density for these three volume-limited samples is shown in Figure 4. Again we find that the $\mathrm{EW}_{\mathrm{H}\alpha}$ has a 
strong dependence on galaxy density for the ``full'' sample (red line in Figure~\ref{fig:density_haew}). For all three volume-limited samples we find that $\mathrm{EW}_{\mathrm{H}\alpha}$ 
decreases from values of $\sim$35\AA\ down to $\sim$20\AA\, (Figure~\ref{fig:density_haew}a), $\sim$10\AA\, (Figure~\ref{fig:density_haew}b) and $\sim$15\AA\, (Figure~\ref{fig:density_haew}c).
The trends are similar across the three different volume 
limited samples and they also appear to be similar to the trends of G\'{o}mez et al. (2003) as shown in Figure 2. The reduction in the median $\mathrm{EW}_{\mathrm{H}\alpha}$ with 
increasing density is highly significant.

Figure 3 shows that the ``star-forming'' sample does not uniformly sample all densities, dominating in lower densities. We expect, then, that the relationship between density and 
$\mathrm{EW}_{\mathrm{H}\alpha}$ will be different for the ``star-forming'' sample compared to the ``full'' sample of galaxies.
The blue line in Figure~\ref{fig:density_haew} represents the relationship between $\mathrm{EW}_{\mathrm{H}\alpha}$ and density when only the ``star-forming'' sample of galaxies is used. The trend 
between $\mathrm{EW}_{\mathrm{H}\alpha}$ and density is essentially absent for the star-forming population.

The selection of star-forming galaxies 
carries further significance when considering SFR as opposed to simply EW and it would be expected that SFR 
will also show weak trends, if any, against density. This is investigated more thoroughly in the following section.
The immediate implication is that the \citet{Gom:03} results are driven by the increasing proportion of the passive galaxy population and not necessarily by a physical suppression driven
by the density of the environment.

This is consistent with the result of McGee et al. (2011) and supports the work of Balogh et al. (2004, 2006) who showed a lack of SFR dependence on the environment
for the star-forming population.




\section{Disentangling SFR, Stellar Mass and Density}

The results of the previous section leave it unclear as to
how much of an effect density has on star-forming galaxies. While a strong relationship between stellar mass and SFR in galaxies is well established (e.g., Peng et al., 2010), 
the above results
question the extent to which a galaxy's local density plays a role in suppressing star-formation in star-forming galaxies. In order to disentangle the three-way relationship between SFR, stellar 
mass and density we investigate the three parameters as functions of each other. This demonstrates the dominant effect of stellar mass, rather than environment, in governing the SFR 
distribution consistent with the results of Peng et al. (2010).

While we only show relationships between $\mathrm{SFR}_{\mathrm{H}\alpha}$ and density and $\mathrm{EW}_{\mathrm{H}\alpha}$ density the same trends are observed with SSFR.
This is a necessary first step, in order to eliminate any biases that might be introduced simply due to the changing population mix with density.
This investigation uses only our ``star-forming'' sample, within each volume-limited sample.

\subsection{As a function of SFR and EW}

\begin{figure*}
\centerline{\includegraphics[width=58mm, height=58mm]{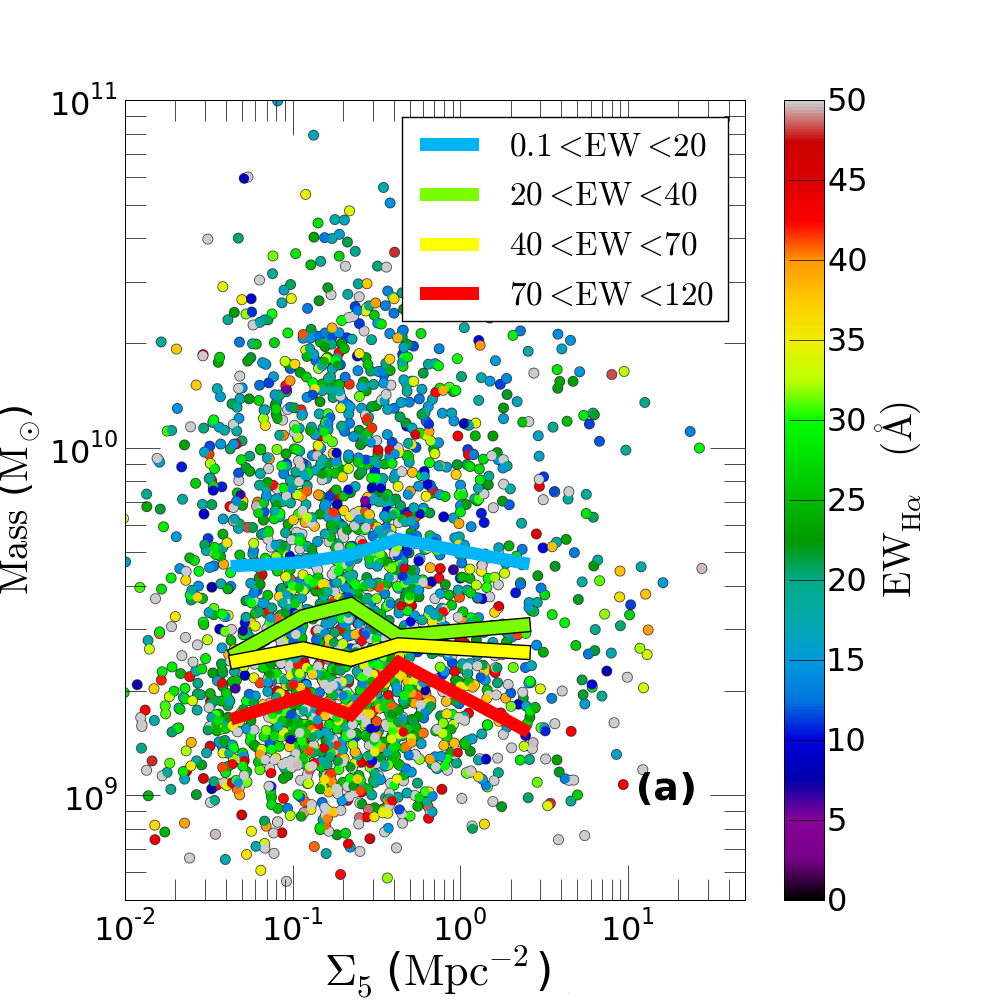}
\includegraphics[width=58mm, height=58mm]{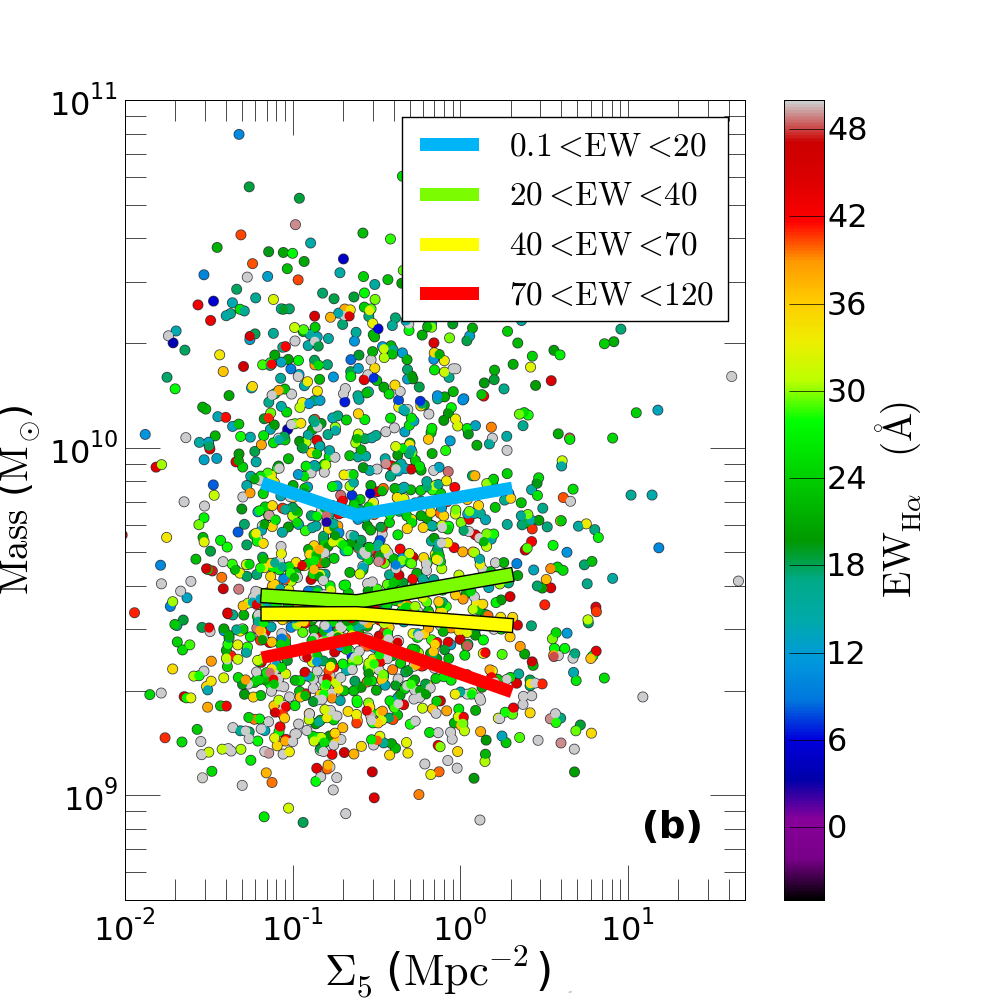}
\includegraphics[width=58mm, height=58mm]{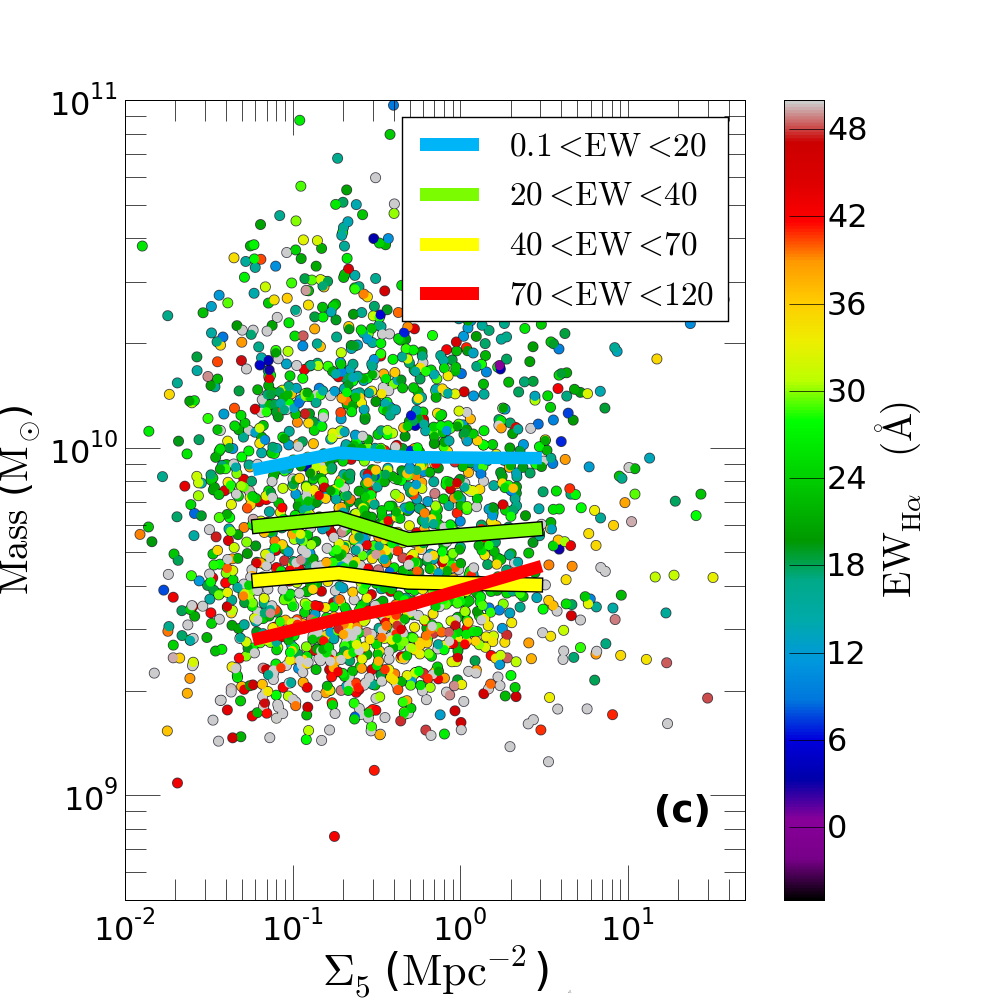}}
\caption{Stellar mass as a function of density with colour representing $\mathrm{EW}_{\mathrm{H}\alpha}$ for VL1, VL2 and VL3 (panels a, b \& c respectively) for the ``star-forming'' sample. 
The lines were derived by taking the median stellar mass in bins of EW, within a series of 
density bins. There is no relationship between mass and density for any $\mathrm{EW}_{\mathrm{H}\alpha}$.}
\label{ew_ew_bins}
\end{figure*}

\begin{figure*}
\centerline{\includegraphics[width=58mm, height=58mm]{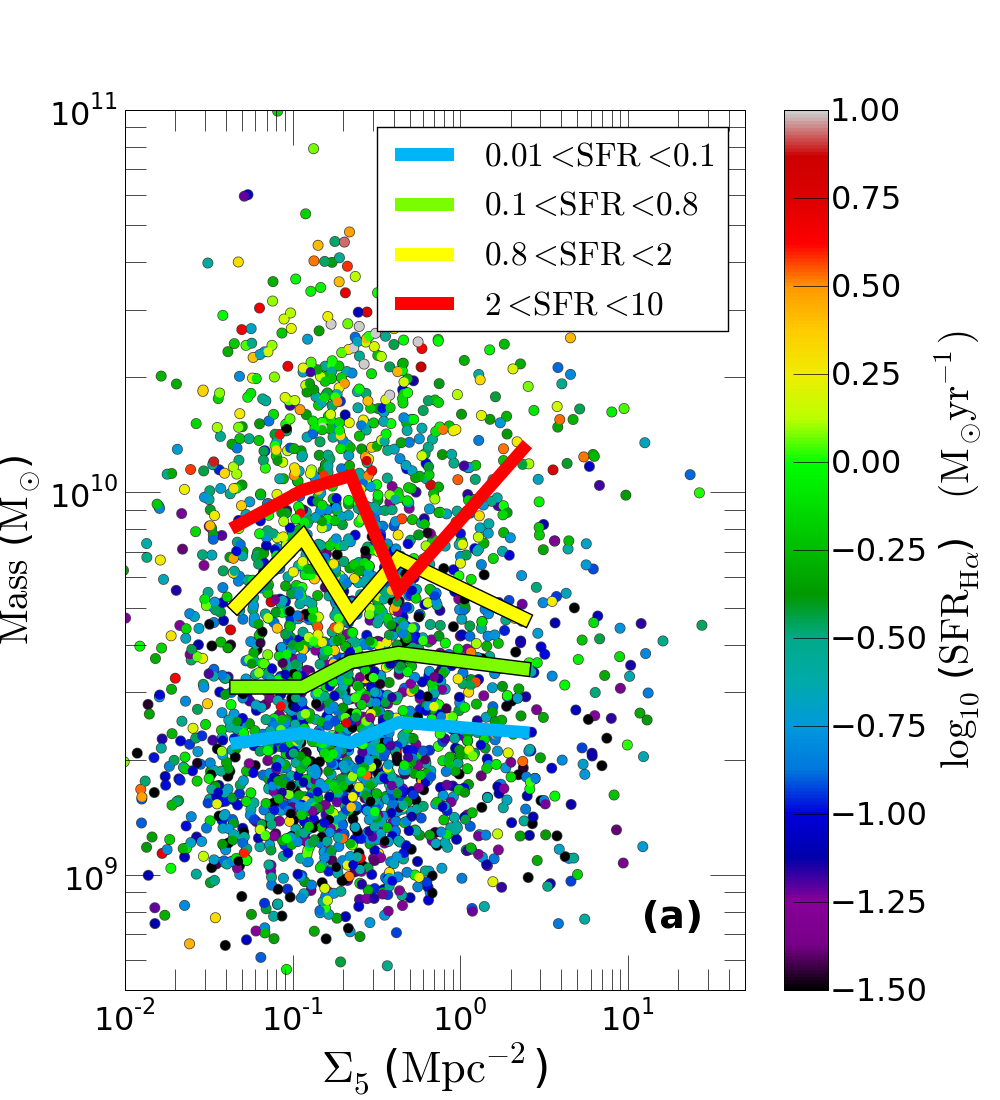}
\includegraphics[width=58mm, height=58mm]{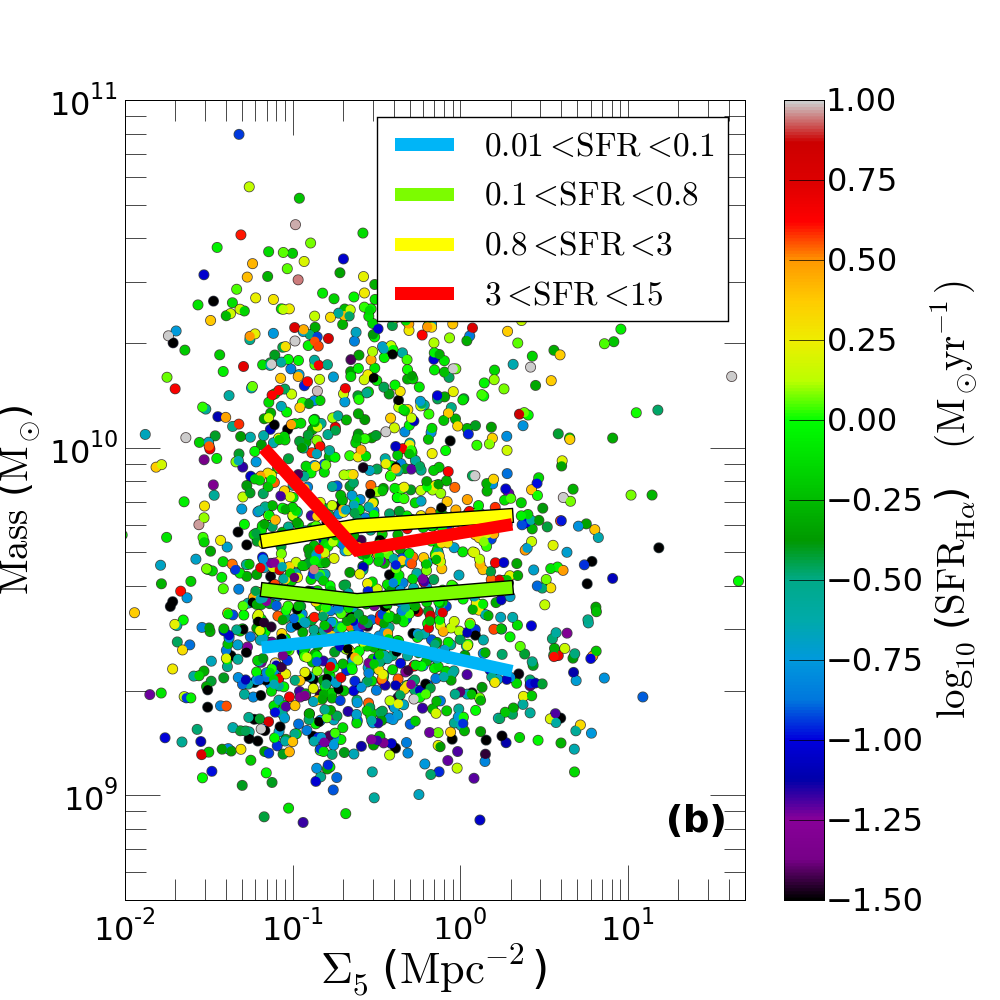}
\includegraphics[width=58mm, height=58mm]{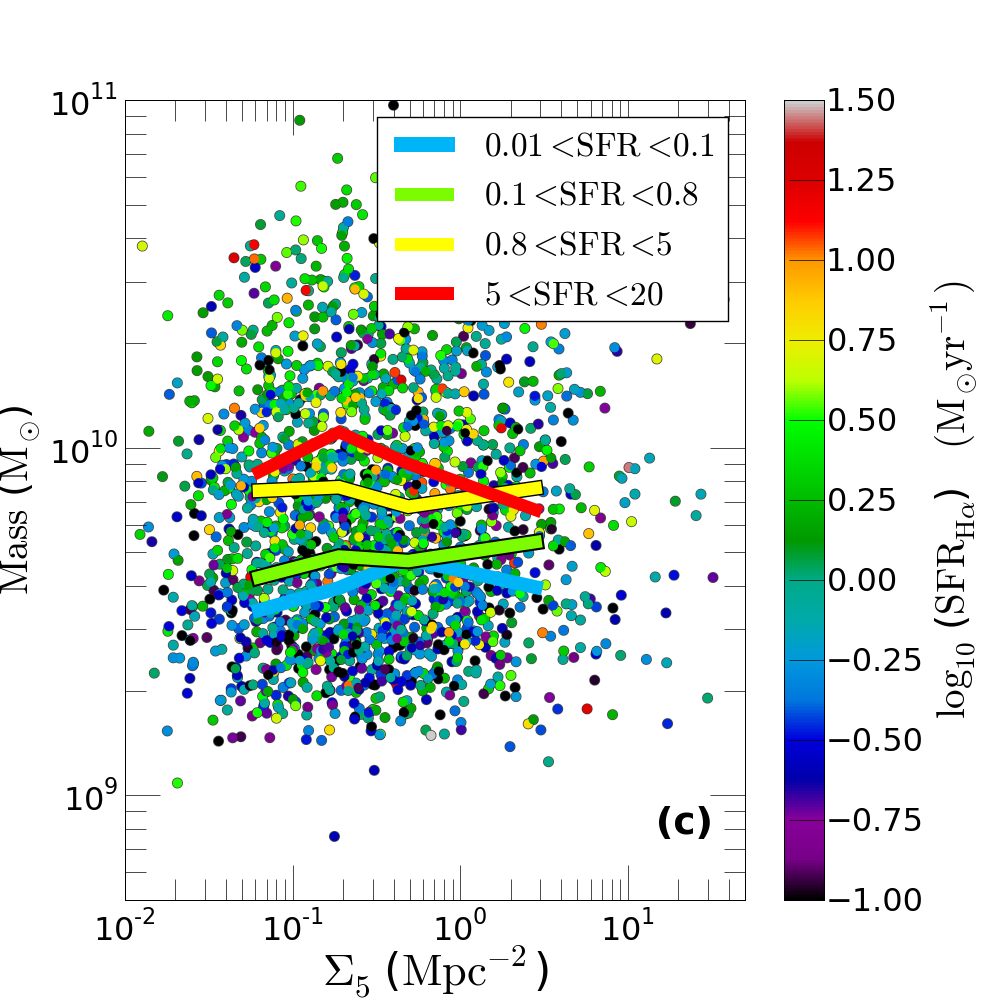}}
\caption{Stellar mass as a function of density with colour representing the H$\alpha$ derived SFRs for VL1, VL2 and VL3 (panels a, b \& c respectively) for the ``star-forming'' sample.
The lines were derived by taking the median 
stellar mass in bins of SFR, within a series of density bins. 
There is no relationship between mass and density for any $\mathrm{SFR}_{\mathrm{H}\alpha}$.}
\label{sfr_sfr_bins}
\end{figure*}

Figures~\ref{ew_ew_bins} and \ref{sfr_sfr_bins} show galaxy stellar mass as a function of density, with medians in bins of $\mathrm{EW}_{\mathrm{H}\alpha}$ (Figure~\ref{ew_ew_bins}) and SFR
(Figure~\ref{sfr_sfr_bins}). These Figures do not show a strong dependence of stellar mass on density for any given EW or SFR.
We do, however, observe that the EWs and SFRs are affected by stellar mass, as
there is a clear separation in the median lines between EW and SFR bins.

The median lines also highlight the absence of a relationship between density and stellar mass for varying SFR and EW. In other words, the high and the low SFR and EW bins show 
similarly flat trends between stellar mass and density.

The sampling of higher masses at higher redshift means that the ranges of SFR and EW shown (in the ``star-forming'' sample) increase systematically with redshift. 
The median lines, though, show consistent results for each volume-limited sample. We conclude that there is no strong relationship between 
the stellar mass in star-forming galaxies and density, as a function either of 
$\mathrm{EW}_{\mathrm{H}\alpha}$ or SFR.

\subsection{As a function of stellar mass}

\begin{figure*}
\centerline{\includegraphics[width=58mm, height=58mm]{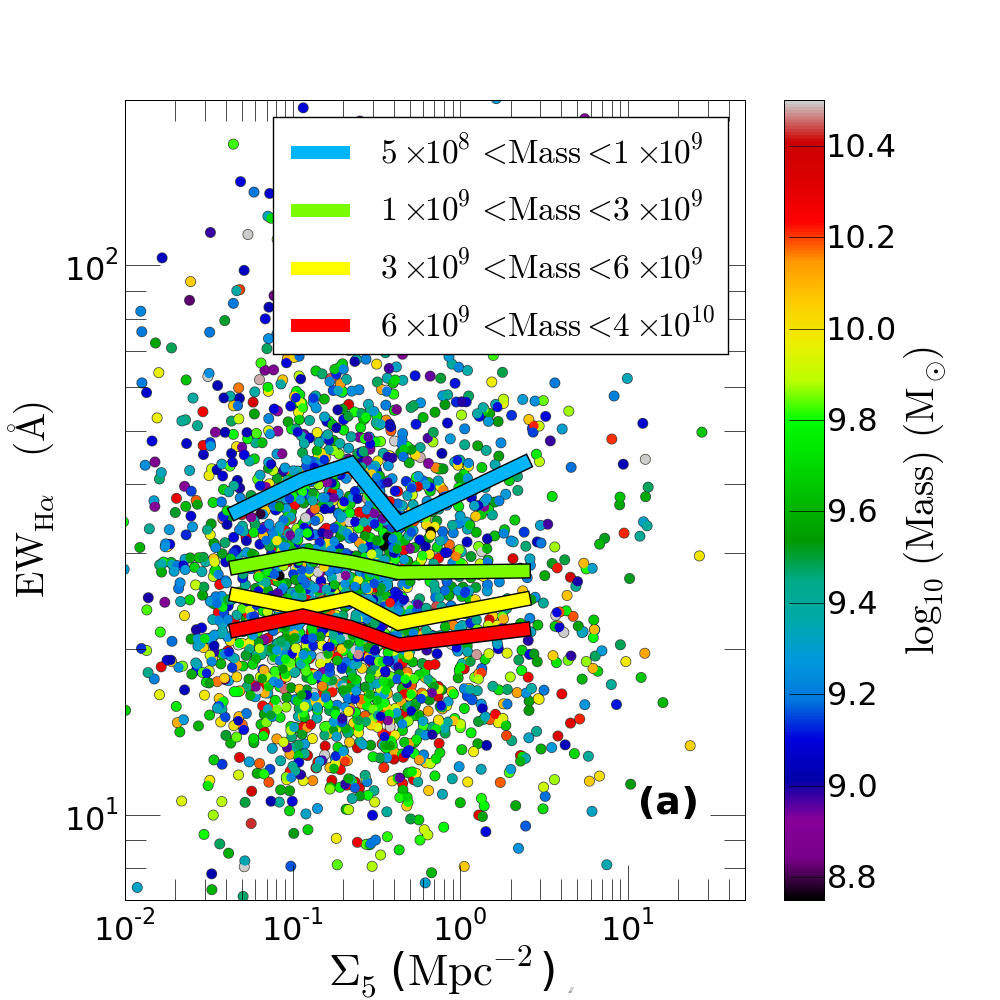}
\includegraphics[width=58mm, height=58mm]{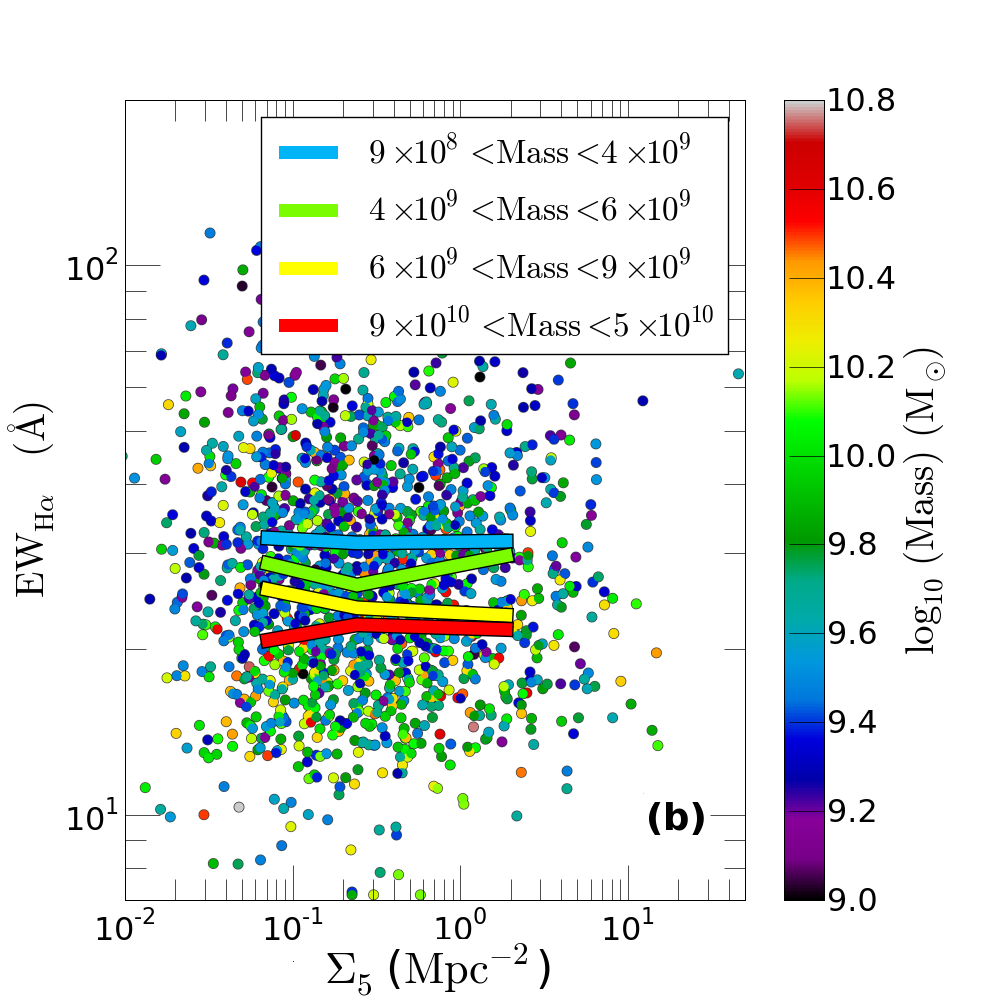}
\includegraphics[width=58mm, height=58mm]{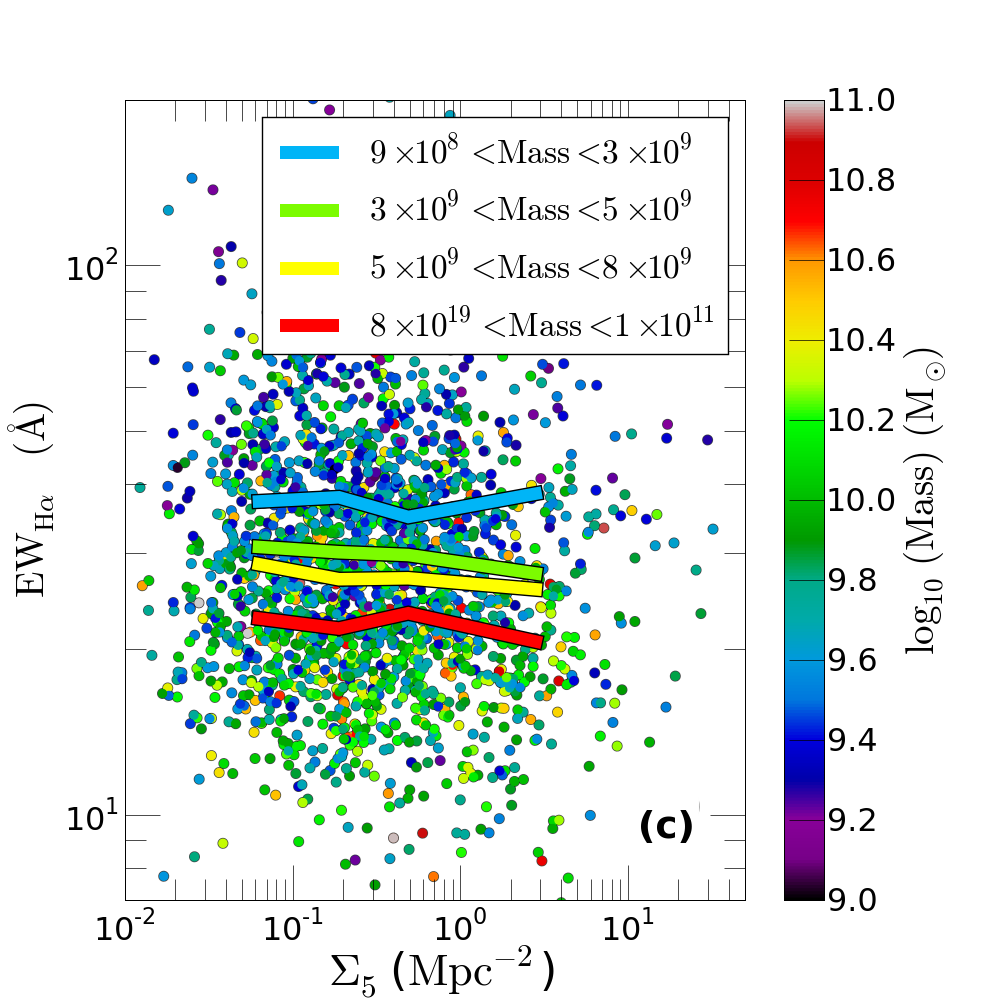}}
\caption{$\mathrm{EW}_{\mathrm{H}\alpha}$ as a function of density with colour representing the stellar mass for VL1, VL2 and VL3 (panels a, b \& c respectively) for the ``star-forming'' sample. 
The lines were derived by taking the median EWs in bins of stellar mass within a series of 
density bins. There is no relationship between $\mathrm{EW}_{\mathrm{H}\alpha}$ and density for any stellar mass.}
\label{ew_mass_bins}
\end{figure*}

\begin{figure*}
\centerline{\includegraphics[width=58mm, height=58mm]{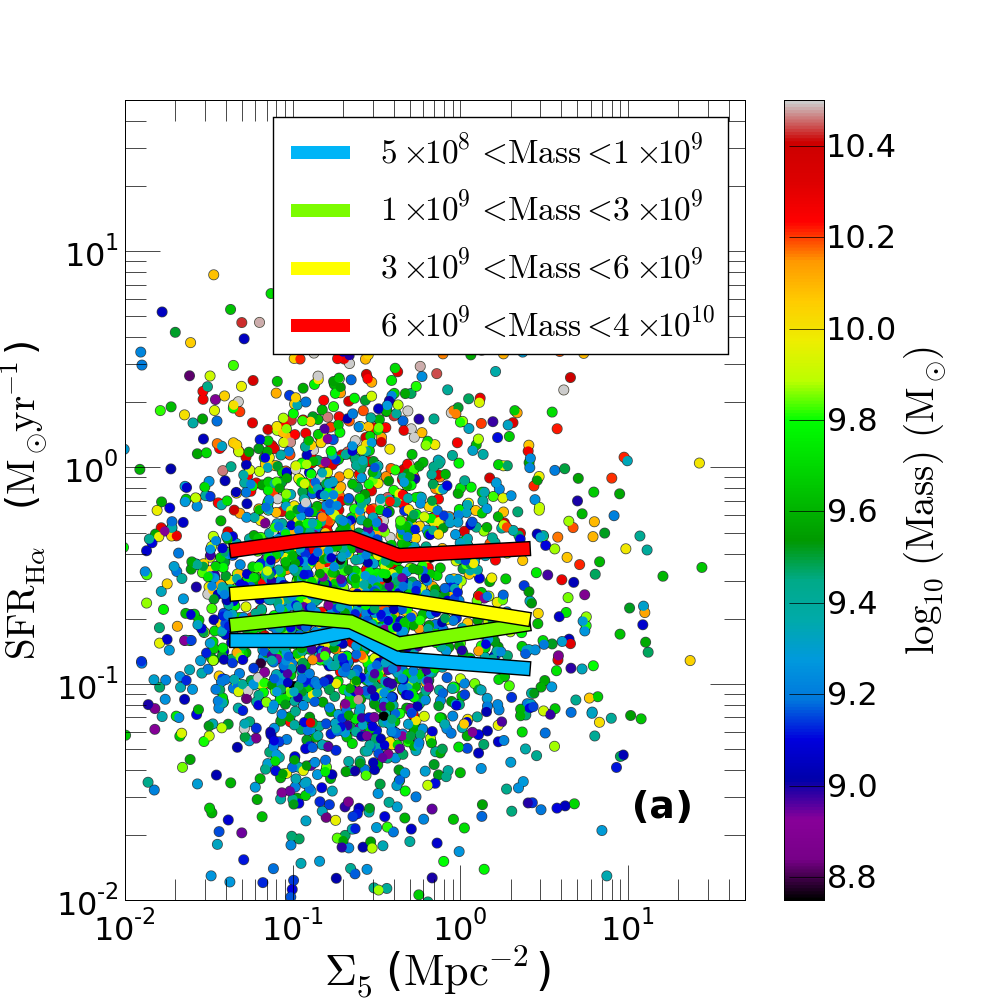}
\includegraphics[width=58mm, height=58mm]{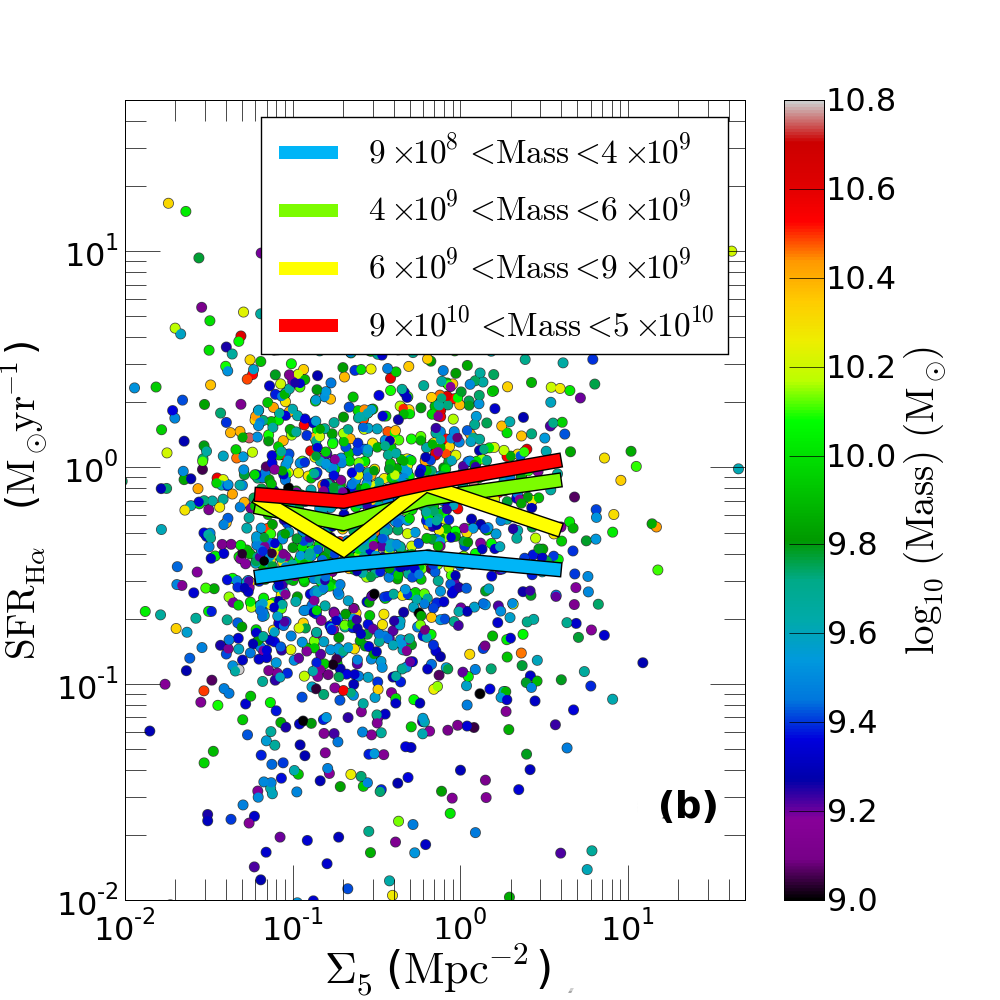}
\includegraphics[width=58mm, height=58mm]{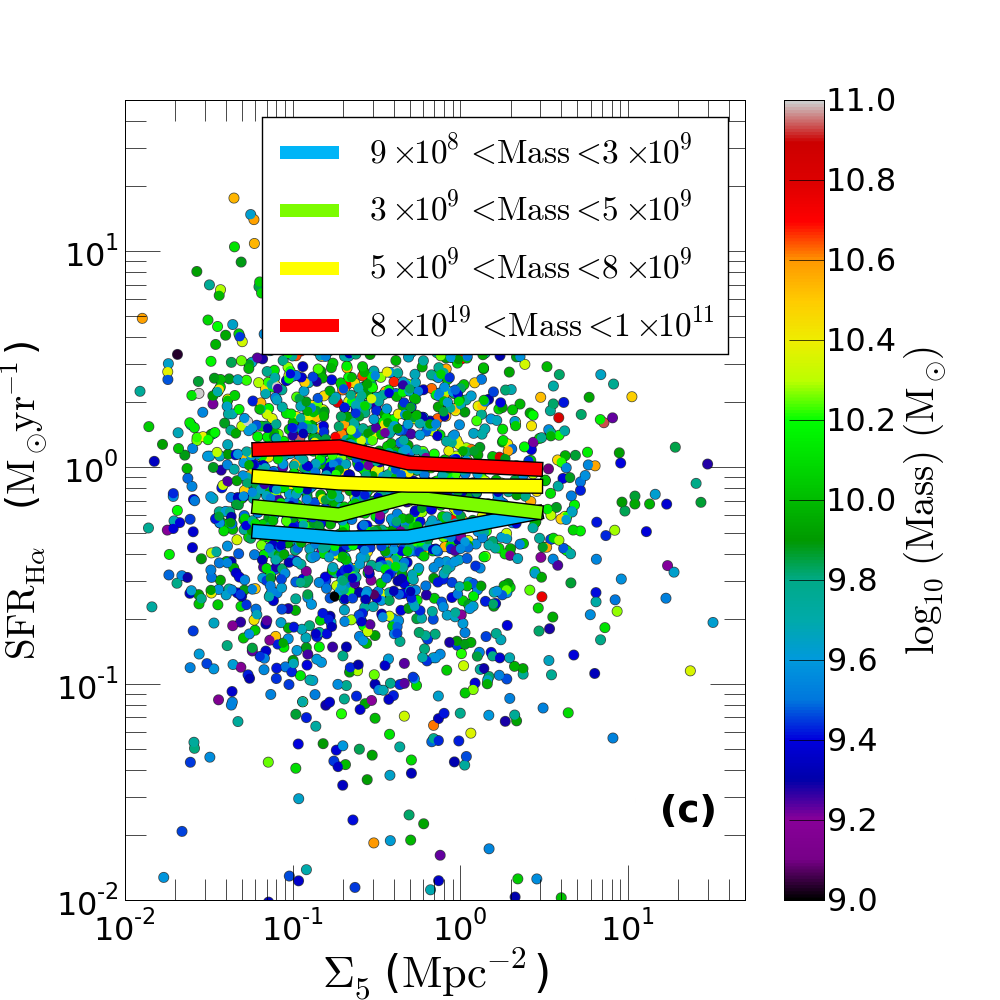}}
\caption{H$\alpha$ derived SFR as a function of density with colour representing stellar mass for VL1, VL2 and VL3 (panels a, b \& c respectively) for the ``star-forming'' sample. 
The lines were derived by taking the median SFR in bins of stellar mass within a series of density bins.
There is no relationship between $\mathrm{SFR}_{\mathrm{H}\alpha}$ and density for any stellar mass.}
\label{sfr_mass_bins}
\end{figure*}

A comparable analysis exploring the relationship between SFR and EW against density as a function of stellar mass is used to investigate any dependencies between SFR and density 
(Figures~\ref{ew_mass_bins} and \ref{sfr_mass_bins}). 
This analysis examines the consistency of the result obtained in Figures~\ref{ew_ew_bins} and \ref{sfr_sfr_bins}.

Strikingly, we see no relationship between EW and density (Figure~\ref{ew_mass_bins}) or SFR and density (Figure~\ref{sfr_mass_bins}) for any given stellar mass. Again the relationship between SFR 
and stellar mass is is clearly visible from the separation of the median SFR lines for different stellar mass bins. The same is true for the EWs. This confirms the strong relationship observed 
between  stellar mass and SFR (and mass and EW) in Figures~\ref{ew_ew_bins} and \ref{sfr_sfr_bins}.
We also observe the expected increase in median SFR and EW for a given stellar mass range as we move to higher redshifts.

\subsection{As a function of density}

The final and next natural step is to investigate the relationship between stellar mass and SFR (and EW) within bins of density. Now that we have shown there is no relationship between density
and SFR (or EW), a comparison between stellar mass and SFR highlights the strongest influence on SFR and EW. This also clarifies the absence 
of a relationship between density and SFR.

\begin{figure*}
\centerline{\includegraphics[width=58mm, height=58mm]{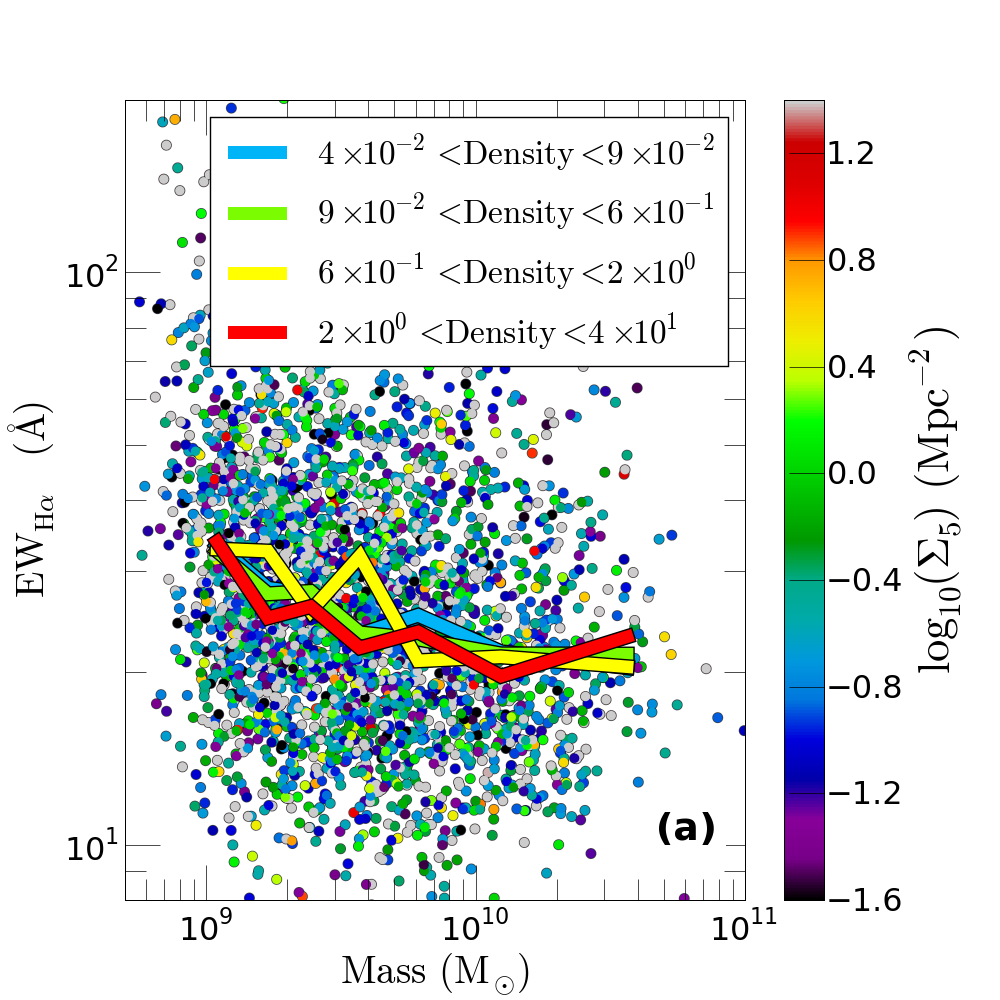}
\includegraphics[width=58mm, height=58mm]{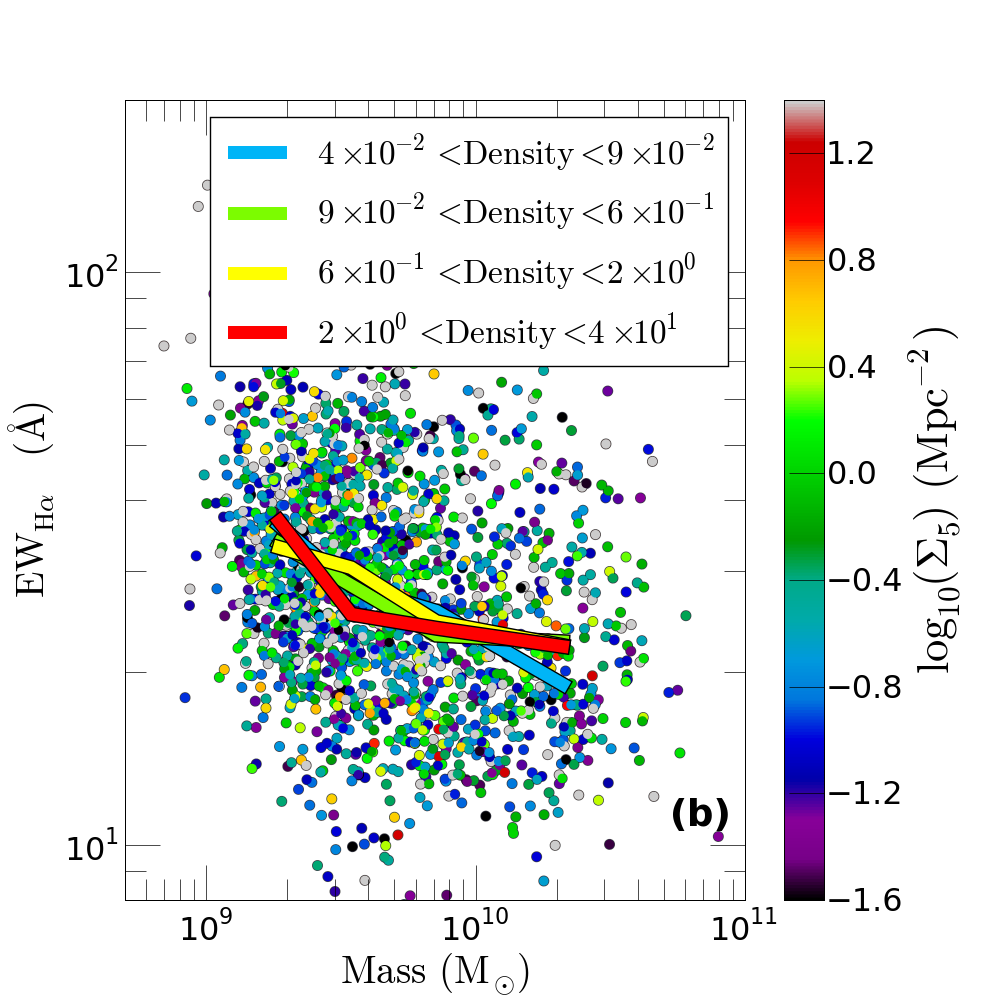}
\includegraphics[width=58mm, height=58mm]{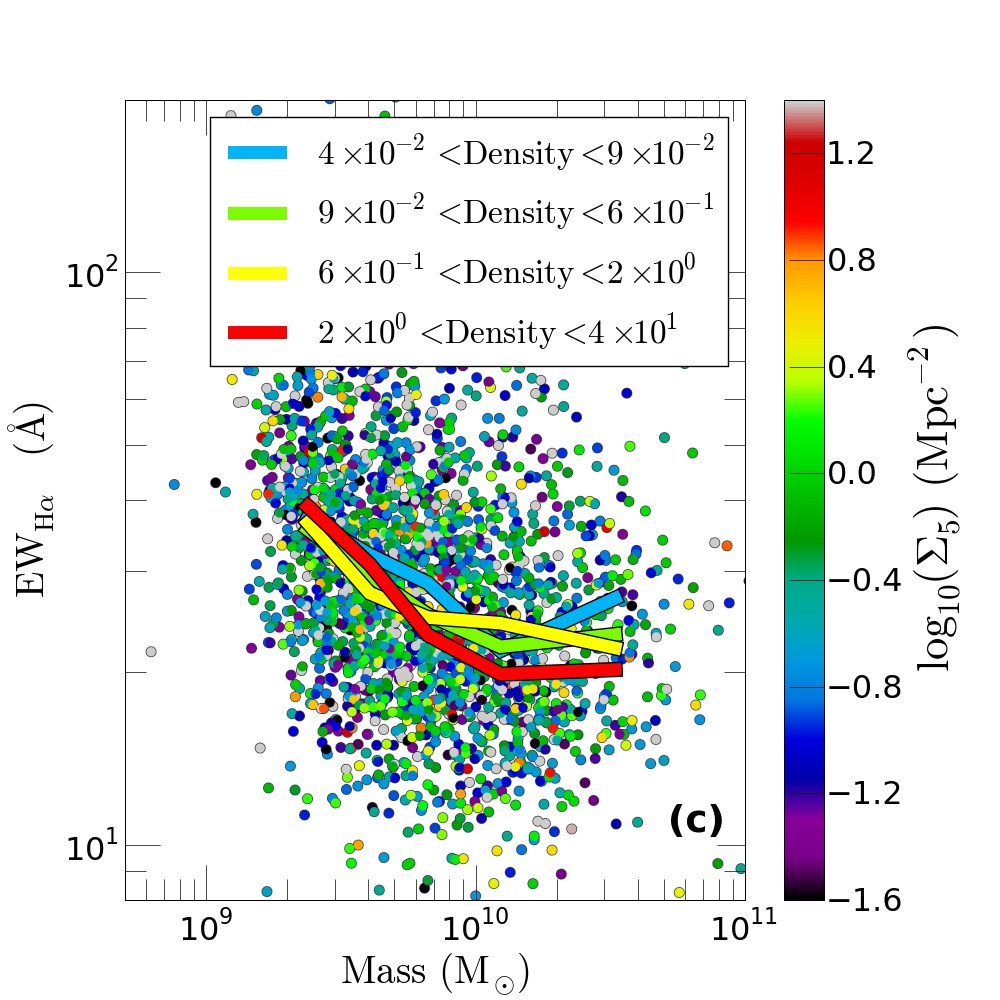}}
\caption{$\mathrm{EW}_{\mathrm{H}\alpha}$ as a function of stellar mass with colour representing density for VL1, VL2 and VL3 (panels a, b \& c respectively) for the ``star-forming'' sample. 
The lines were derived by taking the median EW in bins of density within a series of stellar 
mass bins. $\mathrm{EW}_{\mathrm{H}\alpha}$ and stellar mass are related while the lack of separation
between any of the median lines indicate that there is no real relationship between $\mathrm{EW}_{\mathrm{H}\alpha}$ and density.}
\label{ew_density_bins}
\end{figure*}

Figure~\ref{ew_density_bins} shows a strong dependence between EW and stellar mass, for all density bins in all volume-limited samples, where EW decreases with stellar mass for
all densities. Figure~\ref{ew_density_bins} shows a reduction in EW from $\sim$40\AA\ down to $\sim$20\AA\, with increasing mass.
There does not appear to be a separation between different density bins, highlighting the lack of relation between EW and density, agreeing with the observations made in
Figure~\ref{ew_mass_bins}.

The same is true for SSFR, for which EW is a proxy. Specifically the SSFR trend with mass is not constant as suggested by Peng et al. (2010), but declines strongly as a function of stellar 
mass (see also Bauer et al., in prep). This difference arises since we are considering volume-limited samples in contrast to the apparent magnitude limited sample of Peng et al. (2010), and we are 
also more sensitive to the lower SFR population.

\begin{figure*}
\centerline{\includegraphics[width=58mm, height=58mm]{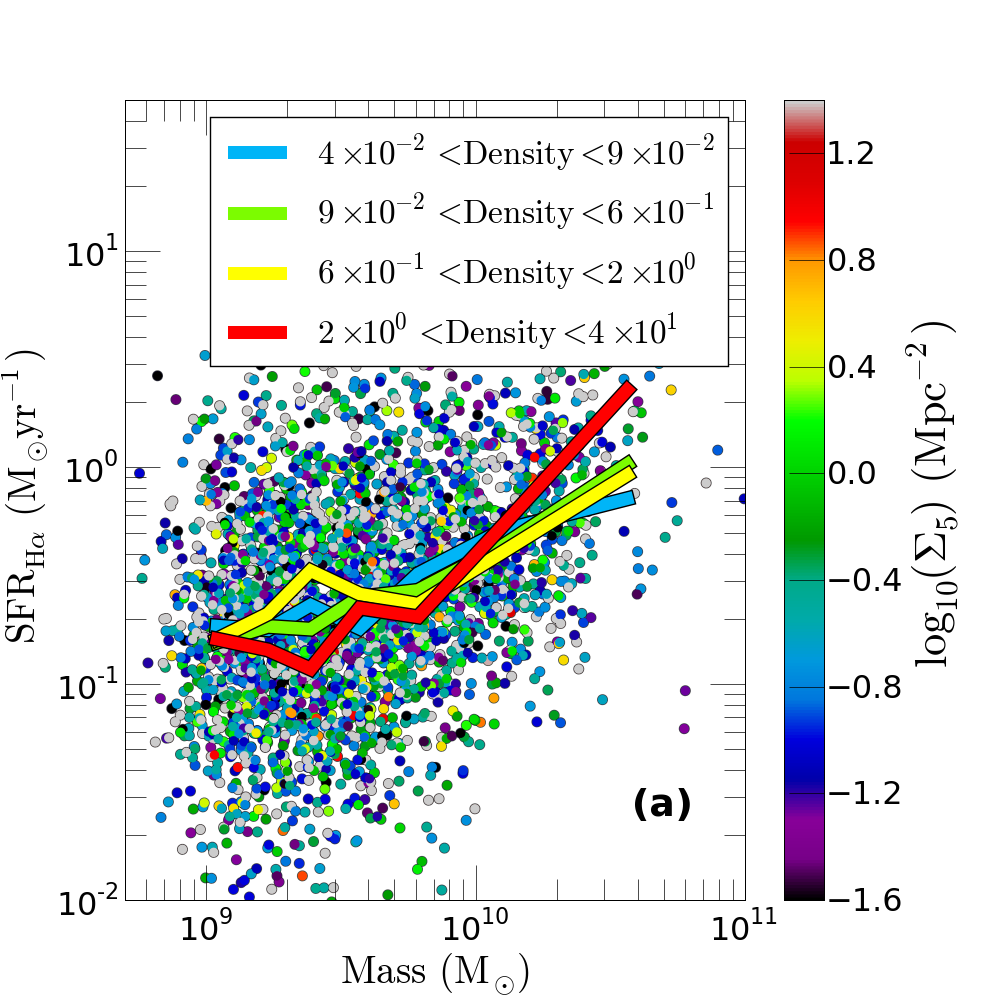}
\includegraphics[width=58mm, height=58mm]{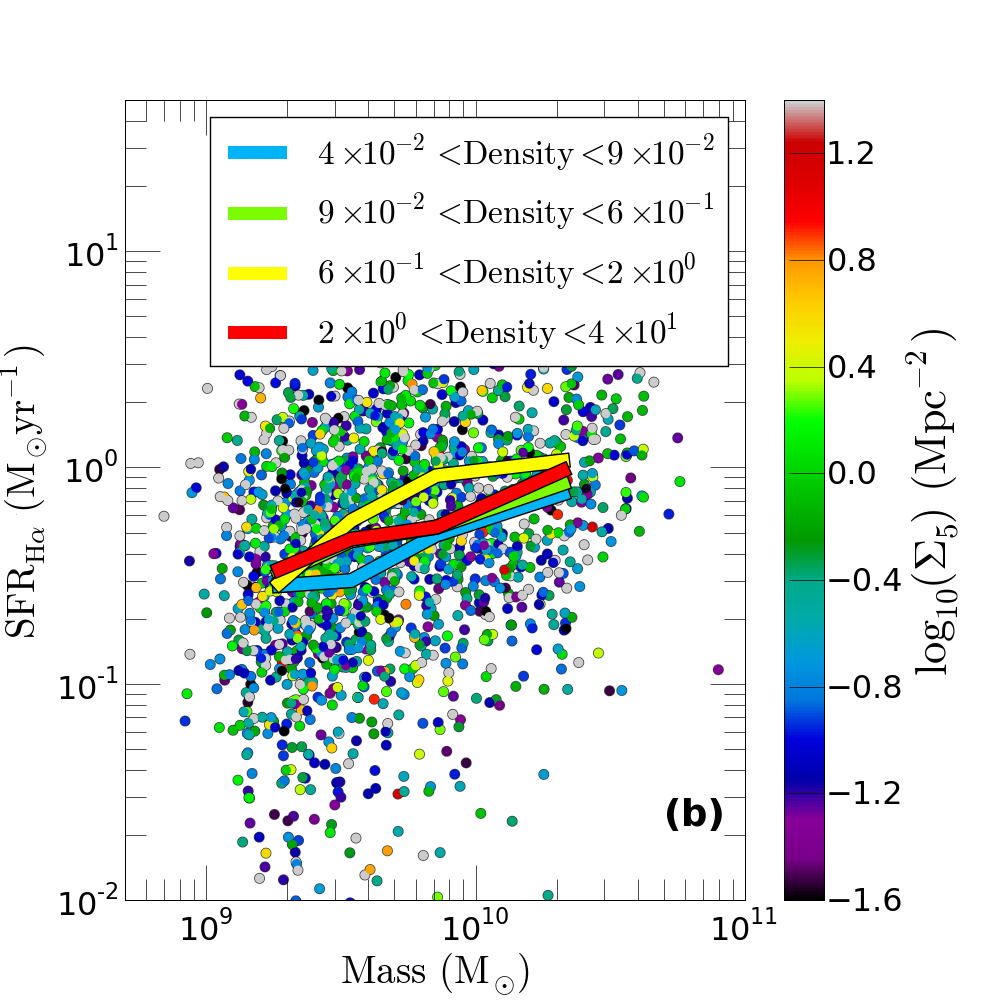}
\includegraphics[width=58mm, height=58mm]{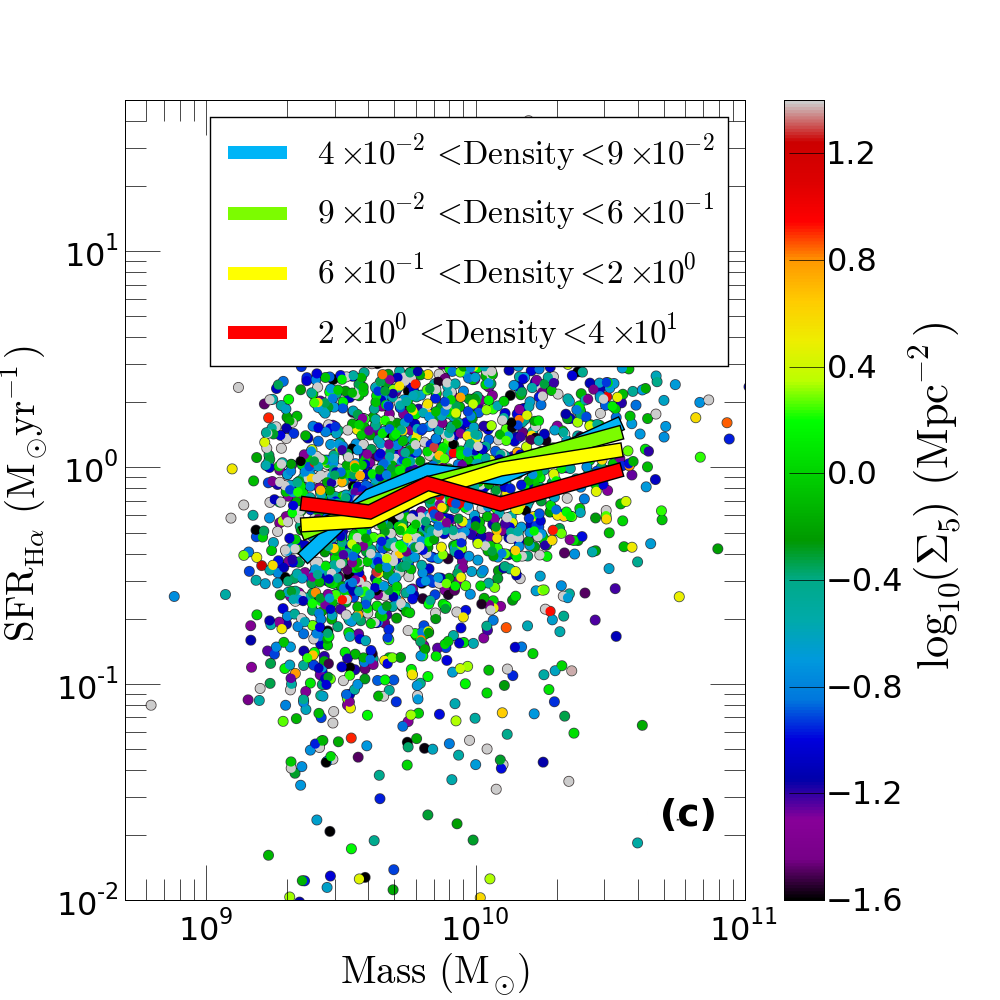}}
\caption{H$\alpha$ derived SFR as a function of stellar mass with colour representing density for VL1, VL2 and VL3 (panels a, b \& c respectively) for the ``star-forming'' sample.
The lines are derived by taking the median SFR in bins of density within a series of stellar mass 
bins. $\mathrm{SFR}_{\mathrm{H}\alpha}$ and stellar mass are related while the lack of separation
between any of the median lines indicate that there is no real relationship between $\mathrm{SFR}_{\mathrm{H}\alpha}$ and density.}
\label{sfr_density_bins}
\end{figure*}

While EW and SSFR show a steady decline with increasing stellar mass, SFR increases with stellar mass (Figure~\ref{sfr_density_bins}).
This observation concurs with the results of Figure~\ref{sfr_mass_bins}.
This figure is similar to Figure 1 of Peng et al. (2010), but demonstrates explicitly our sensitivity to lower SFR systems at a given mass, and the broadening distribution of SFR for lower
stellar mass systems. As with EW, there does not appear to be a significant separation 
between the median SFR lines for each density bin. This indicates a relationship between stellar mass and SFR, but no significant relationship between SFR and density, as well as 
illustrating the absence of an EW-density relationship for the star-forming population.

Figures~\ref{ew_density_bins} and \ref{sfr_density_bins} clearly show that SFR and stellar mass are closely linked and that density appears to have a minimal impact on either of these 
properties, if any, for star-forming galaxies. It appears the $\mathrm{EW}_{\mathrm{H}\alpha}$-density result of G\'{o}mez et al. (2003) is a consequence of the passive galaxy population and is 
not observed in the star-forming population. Likewise, the SFR-density relation of the G\'{o}mez et al. (2003) analysis is also a result of the SFRs that are estimated for the whole population, 
including those for the passive systems. The significance of this is investigated in the next section, where the distribution of 
SFR is analysed as a function of density for the star-forming population.

\section{SFR-Density Relation}

\begin{figure*}
\centerline{\includegraphics[width=58mm, height=58mm]{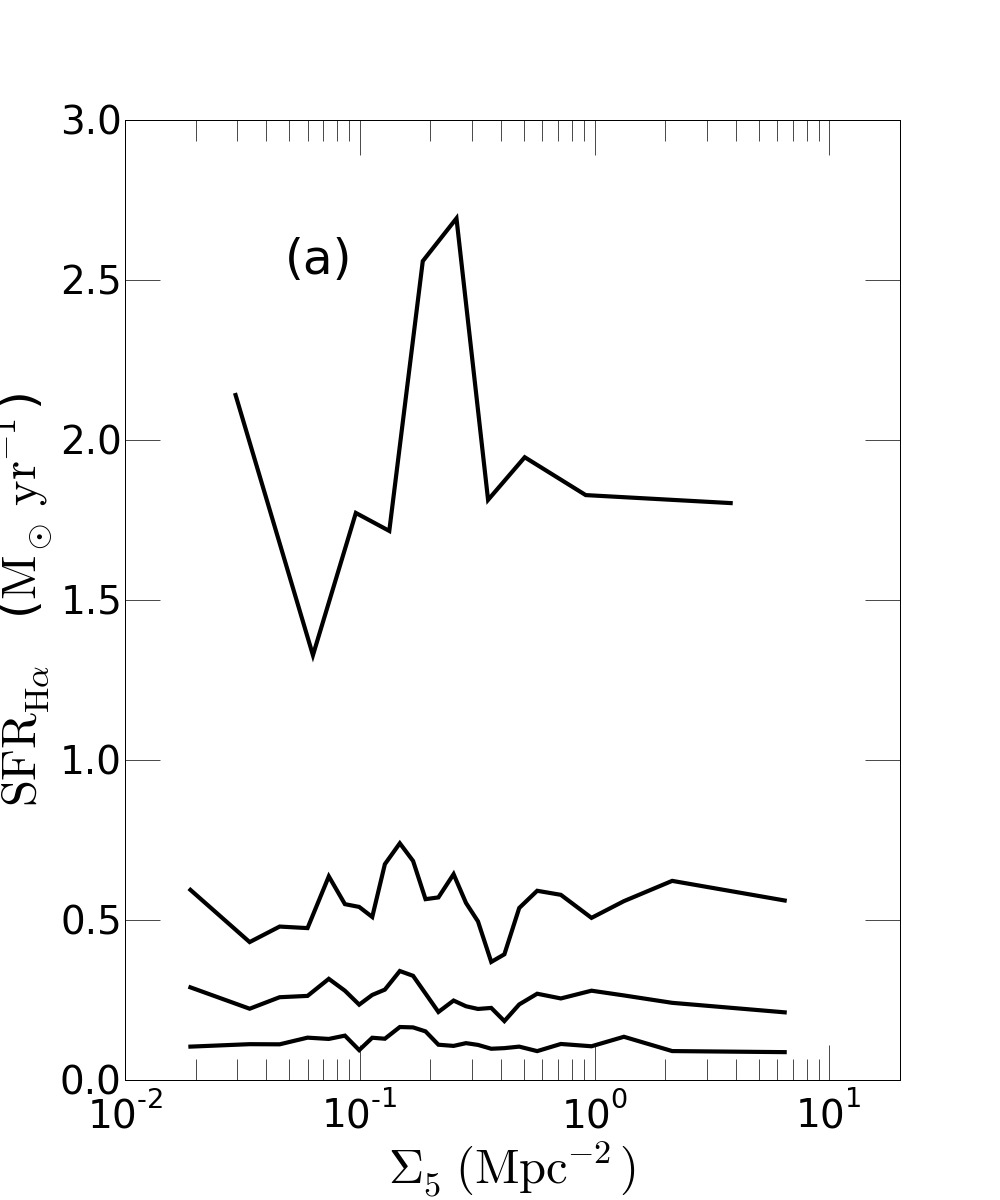}
\includegraphics[width=58mm, height=58mm]{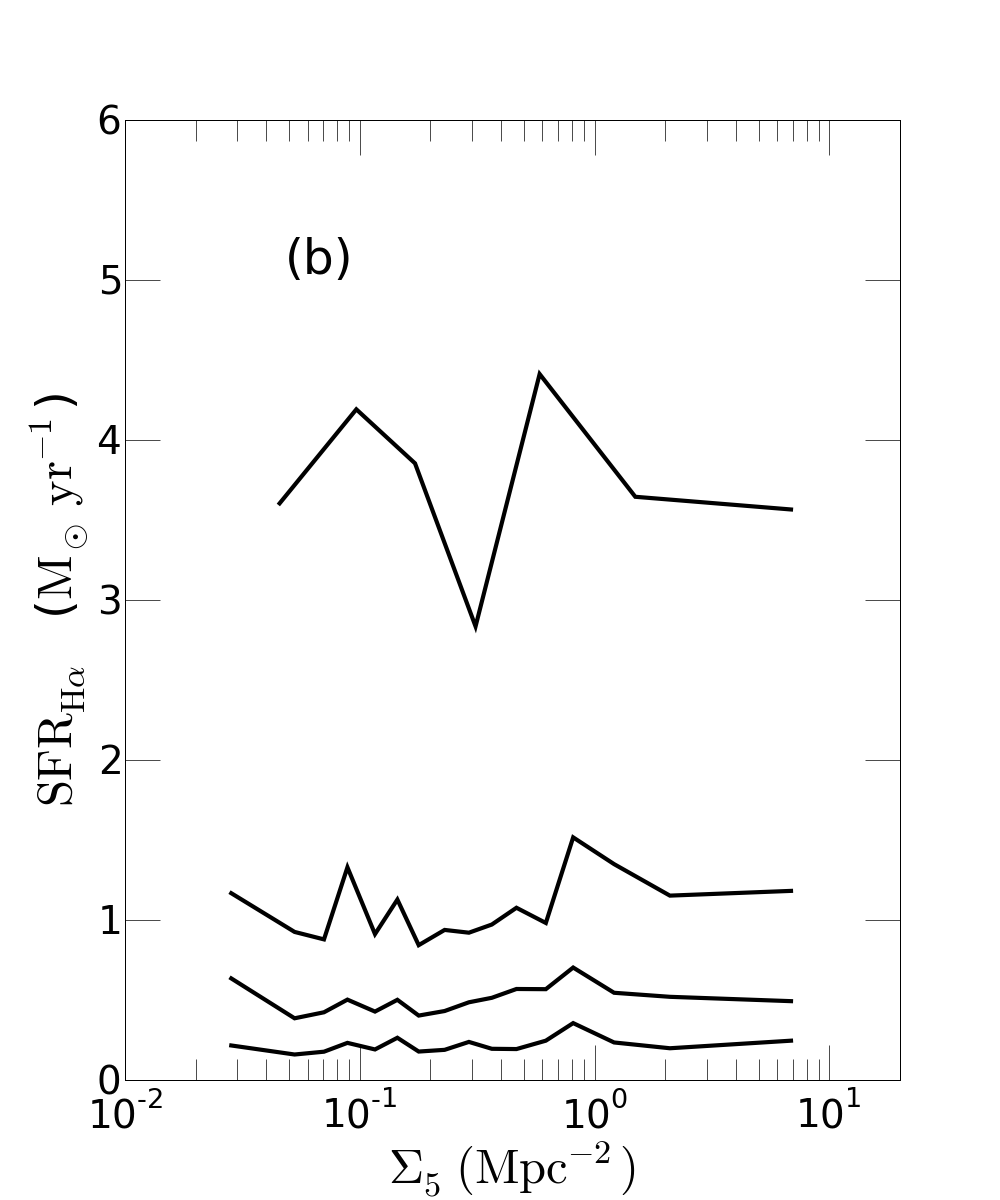}
\includegraphics[width=58mm, height=58mm]{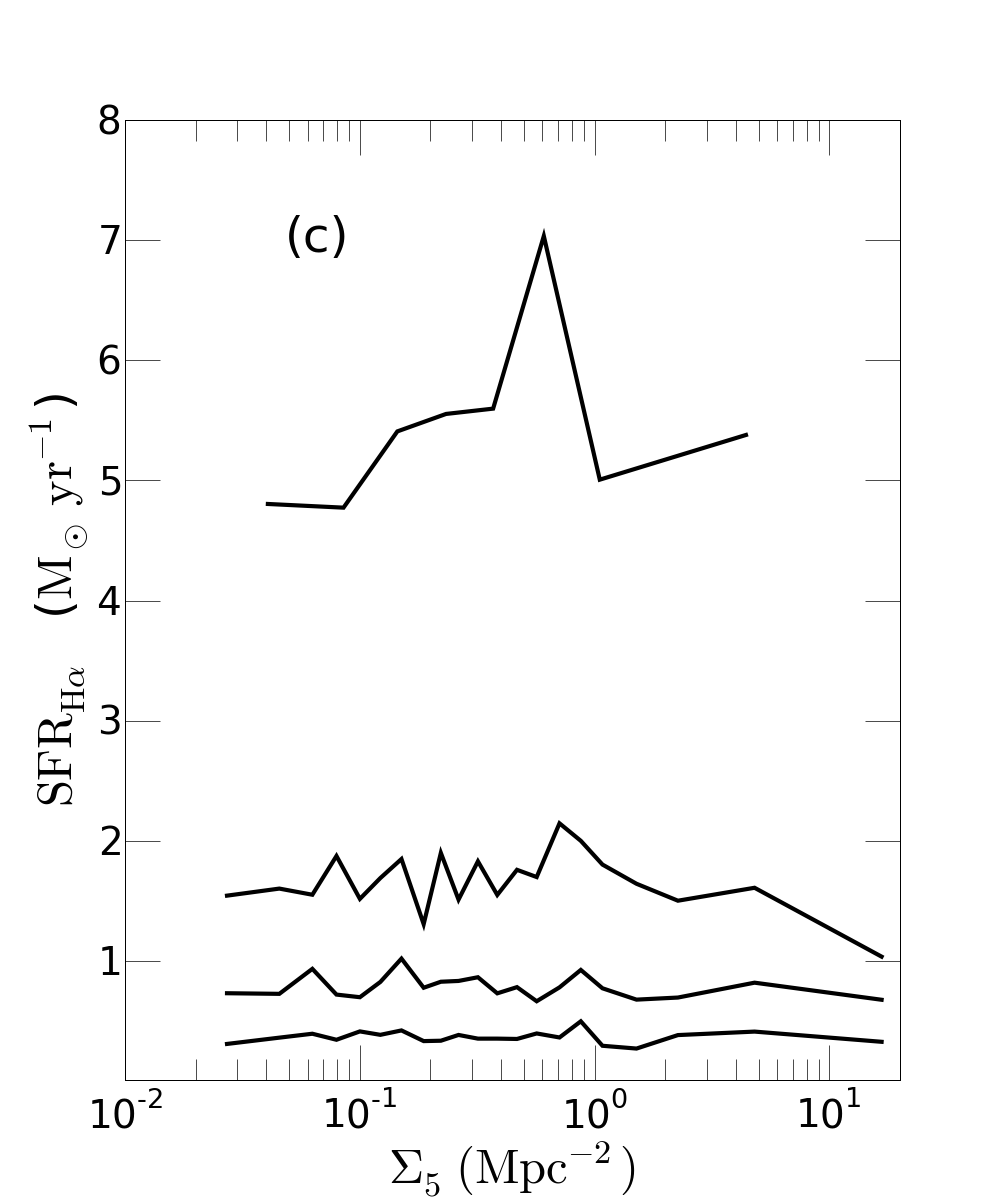}}
\caption{The SFR-density relation for the ``star-forming'' sample of galaxies for the three volume-limited samples. From bottom to top the lines represent the $25^{\mathrm{th}}$, median, 
$75^{\mathrm{th}}$ and $95^{\mathrm{th}}$ percentiles. The density bins each contain at least 100 galaxies.
The distribution of SFR does not change with density, at any redshift.}
\label{fig:density_sfr_ha}
\end{figure*}

We further analyze SFR as a function of density by quantifying the SFR distribution, using a range of percentiles.
The density bins are constructed to ensure that each one contains at least 100 galaxies. If the final bin contains more than 20 galaxies it is considered to be an 
independent bin otherwise these galaxies are included in the preceding bin.
For the calculation of the $95^{\mathrm{th}}$ percentile we use bins of 250 galaxies with a minimum of 50 galaxies to form an independent bin for the final density bin.
This is done to increase the reliability of this high percentile.
We calculate $25^{\mathrm{th}}$, median, $75^{\mathrm{th}}$ and $95^{\mathrm{th}}$ percentiles for SFR for the
three independent volume-limited samples (Figure~\ref{fig:density_sfr_ha}) for the ``star-forming'' sample of galaxies.



\begin{figure*}
\centerline{\includegraphics[width=58mm, height=58mm]{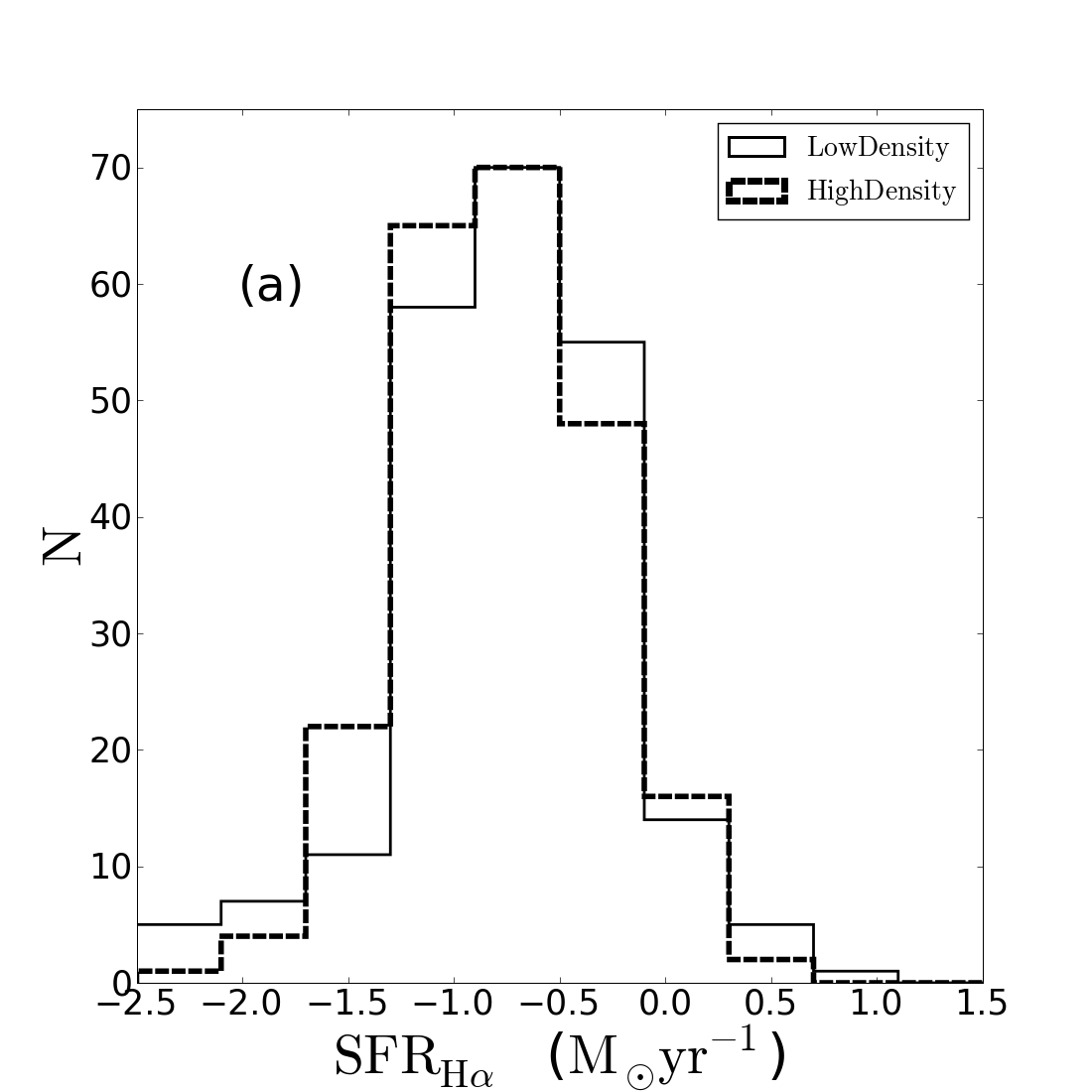}
\includegraphics[width=58mm, height=58mm]{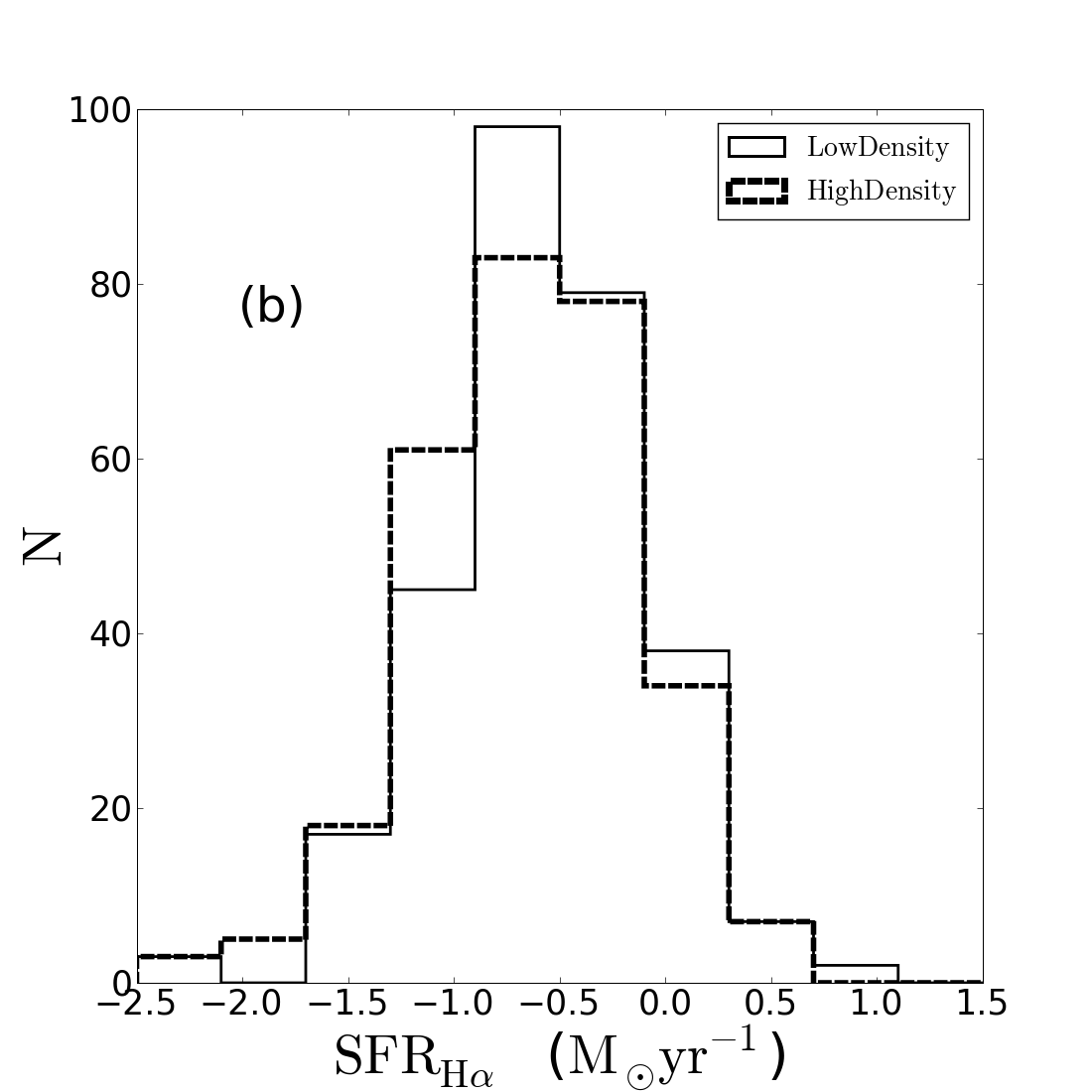}
\includegraphics[width=58mm, height=58mm]{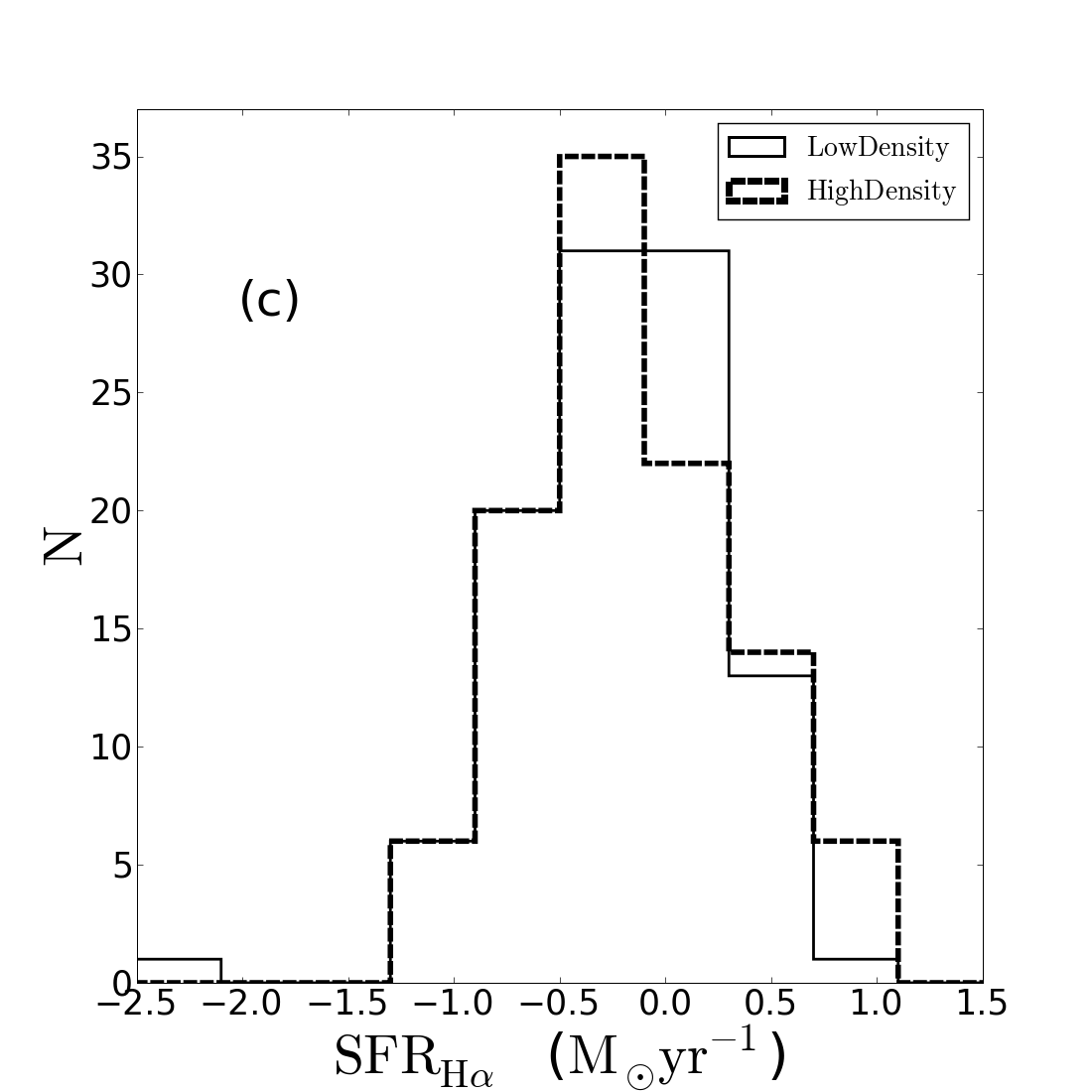}}
\caption{Histograms of SFR for (a) low, (b) mid and (c) high mass bins. The solid line represents the lowest density quartile and the dashed line represents the highest density 
quartile. The distribution of SFRs are similar for both low and high densities for each mass bin. These results are derived from VL1.}
\label{sfr_hist}
\end{figure*}

The change in the distribution of SFR is not nearly as significant as that observed for $\mathrm{EW}_{\mathrm{H}\alpha}$ in the ``full'' sample (Figure~\ref{fig:density_sfr_ha}).
The SFR distribution, as quantified by these percentiles, remains unchanged even in the densest environments.
Figure~\ref{fig:density_sfr_ha} shows no decline even in the $95^{\mathrm{th}}$ percentile, where we sample the most extreme star-forming systems.

The change in the y-axis between the different panels in Figure~\ref{fig:density_sfr_ha} is mostly due to a selection effect: we are more likely to observe brighter, and hence more highly 
star-forming, galaxies at higher redshift. This is seen explicitly in the range of SFRs spanned by the different panels in Figure~\ref{fig:density_sfr_ha}.
The distribution of the SFRs shows no change as a function of density, however, at any redshift explored here.

The lack of a relationship between SFR and density here implies that if there is indeed suppression in galaxy SFRs in high density environments, it 
occurs too rapidly to be observed in our sample as a contraction in the distribution of SFRs for actively star-forming galaxies. Alternatively, galaxies may evolve predominantly 
{\em in-situ\/}, with their evolution being dominated by their mass, and any local environment being a second-order effect.

From the work of Wolf et al. (2009), Vulcani et al. (2010) Li et al. (2011) and \citet{vdL:10} we understand that there may be low SFR galaxies existing in cluster environments that may not 
be seen in the field.
We explore this further in Figure~\ref{sfr_hist} where we compare the SFR distributions in both the lowest and highest density quartiles for a range of mass bins.

We find that all three mass bins show that there is no difference in the SFR distribution for the high and low density bins. 
Kolmogorov-Smirnov (KS) tests carried out on this sample confirm this result with, low, mid and high mass bins 
showing KS statistics of 0.535, 0.307, and 0.576. These values need to be below 0.05 for the two density distributions to be classified as being selected from two different galaxy populations. While
the results of Figure~\ref{sfr_hist} are derived for one volume-limited sample (VL1) the other two volume-limited samples show similar results.

\section{Morphology and Density}

Morphology and density are related in the local universe such that early-type galaxies
tend to dominate in dense environments, such as the center
of clusters, and late-types are more prominent in low-density environments \citep[e.g.,][]{Drs:80}. G\'{o}mez et al. (2003), using concentration index as a proxy for morphology, claim that the 
SFR-density relationship is not simply a
consequence of this well-established morphology-density relationship, but that it holds within a given morphological type. We explore that result in more detail here. 

We use the analysis of \citet{Ltz:04} and three morphological parameters, the concentration (C) \citep{Con:03}, Gini coefficient (G) and the brightest 20\% of the 2nd moment 
of the flux ($\mathrm{M}_{20}$), to classify our samples into early and late-types.

Lotz et al. (2004) showed that galaxies with low concentration and Gini coefficient values and high $M_{20}$ values can be deemed to be late-type galaxies to a high probability. 
Galaxies with high concentration and Gini coefficient values and low $M_{20}$ values can be grouped as early-types. The values used to make these classifications are given in Table 4.
It must be noted that these classifications are conservative, in that only galaxies obeying all criteria were classified as late or early, with many galaxies remaining unclassified.
In this manner we can be confident that galaxies we call early or late are robustly classified as such.

\begin{table}
\centering
\caption{Morphology classification parameters. Galaxies were classified as early or late only if they obeyed all criteria, otherwise they remain unclassified.}
\begin{tabular}{cc} 
\hline\hline \\
Early-type & Late-type \\ [0.5ex]
\hline \\
   $C>2.8$ & $C<2.75$ \\
$G>0.55$ & $G<0.65$ \\
$M_{20} < -1.65$ & $M_{20} > -1.6$\\[1ex]
\hline
\end{tabular}
\label{table:morph}
\end{table}

\begin{figure*}
\centerline{\includegraphics[width=58mm, height=58mm]{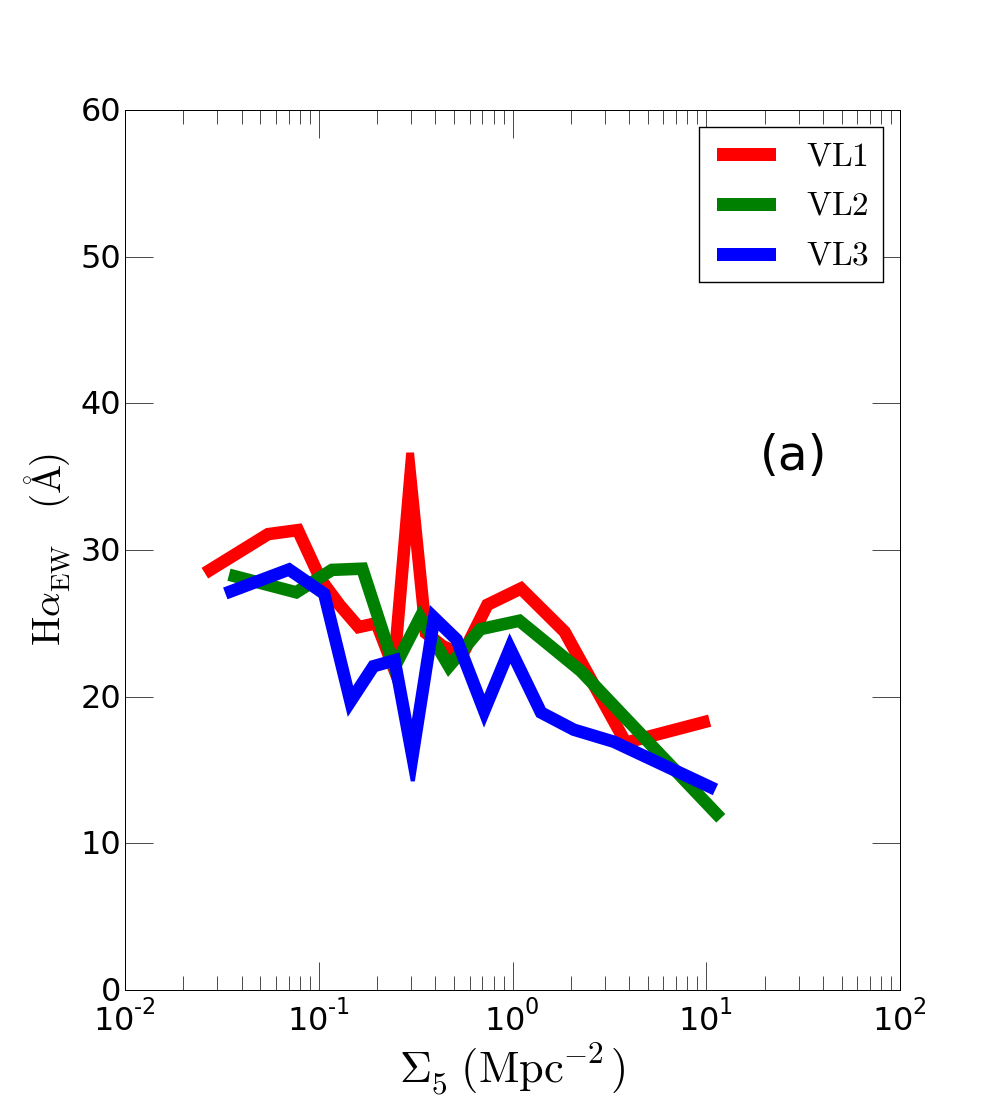}
\includegraphics[width=58mm, height=58mm]{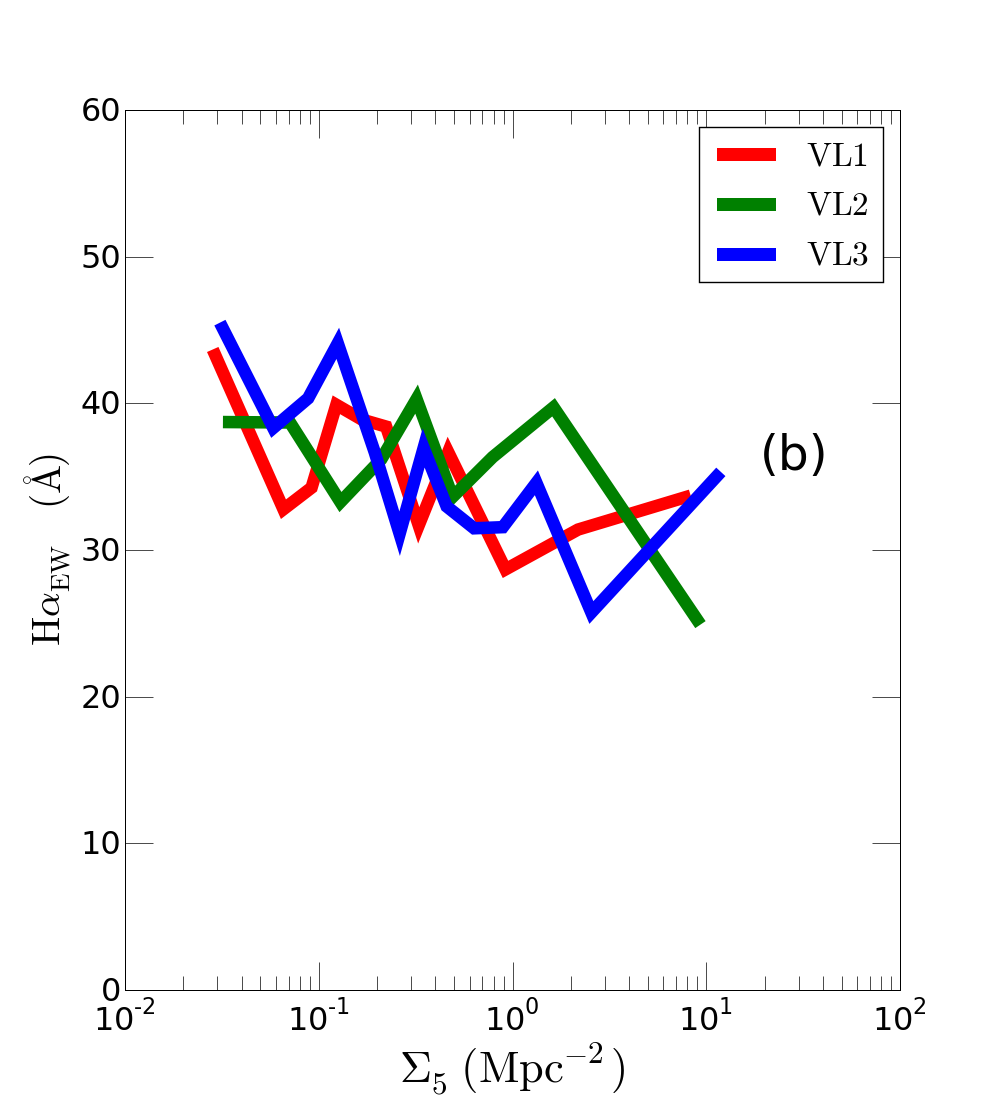}}
\caption{$\mathrm{EW}_{\mathrm{H}\alpha}$  as a function of density for the ``full'' sample of galaxies for (a) early-types and (b) late-types for the three volume-limited samples (red: VL1, 
green: VL2, blue: VL3). The lines represent the $75^{\mathrm{th}}$ percentile. Panel (a) uses 100 galaxies per bin while
panel (b) uses 50 galaxies per bin due to the smaller numbers of galaxies in this sample. Early-type galaxies show a strong decrease in $\mathrm{EW}_{\mathrm{H}\alpha}$ with increasing 
density for the ``full sample'' of galaxies. Late-types show a similar but a much weaker trend for the ``full sample'' of galaxies.}
\label{fig:full}
\end{figure*}

We again look for variations between density and $\mathrm{EW}_{\mathrm{H}\alpha}$ in both the full and ``star-forming'' samples using the 75$^{\mathrm{th}}$ percentile, now looking at early and 
late-type galaxies. Figure~\ref{fig:full} shows the results for the ``full'' sample for which we have 2909, 1885 and 2604 early-type galaxies and 723, 587 and 847 late-types for 
VL1, VL2 and VL3 respectively. For the ``star-forming'' sample there are 1089, 520 and 595 early-type galaxies and 545, 375 and 514 late-type galaxies.
Early-type galaxies in the ``full'' sample show a strong relationship between $\mathrm{EW}_{\mathrm{H}\alpha}$ and density.
This contrasts with the equivalent relationship for the late-type galaxies, which has a much weaker trend between density and EW. This is true for all three volume-limited samples.

Furthermore, we see that it is the early-type galaxies that drive the morphology density relation for the ``full'' sample of galaxies with perhaps a minor contribution from the late-type galaxies.
This is supported by the fact that at the highest densities, the median EW of early-type galaxies is near 10\AA\ while late-types reach a minimum EW of about 30\AA\ in the highest density bin.
The reduction of EW in early-types for the ``full'' samples is as much as 70\% while the late-types experience a maximum decrease in EW of $\sim$25\%.
The fact that the EW for the entire sample of galaxies (including early and late-types as in Figure 4) also approaches very low $\mathrm{EW}_{\mathrm{H}\alpha}$ in the highest density bins 
demonstrates that the early-types are the main driver behind the reduction in both the $\mathrm{EW}_{\mathrm{H}\alpha}$ as well as the SFR for the full samples as a function of density.

\begin{figure*}
\centerline{\includegraphics[width=58mm, height=58mm]{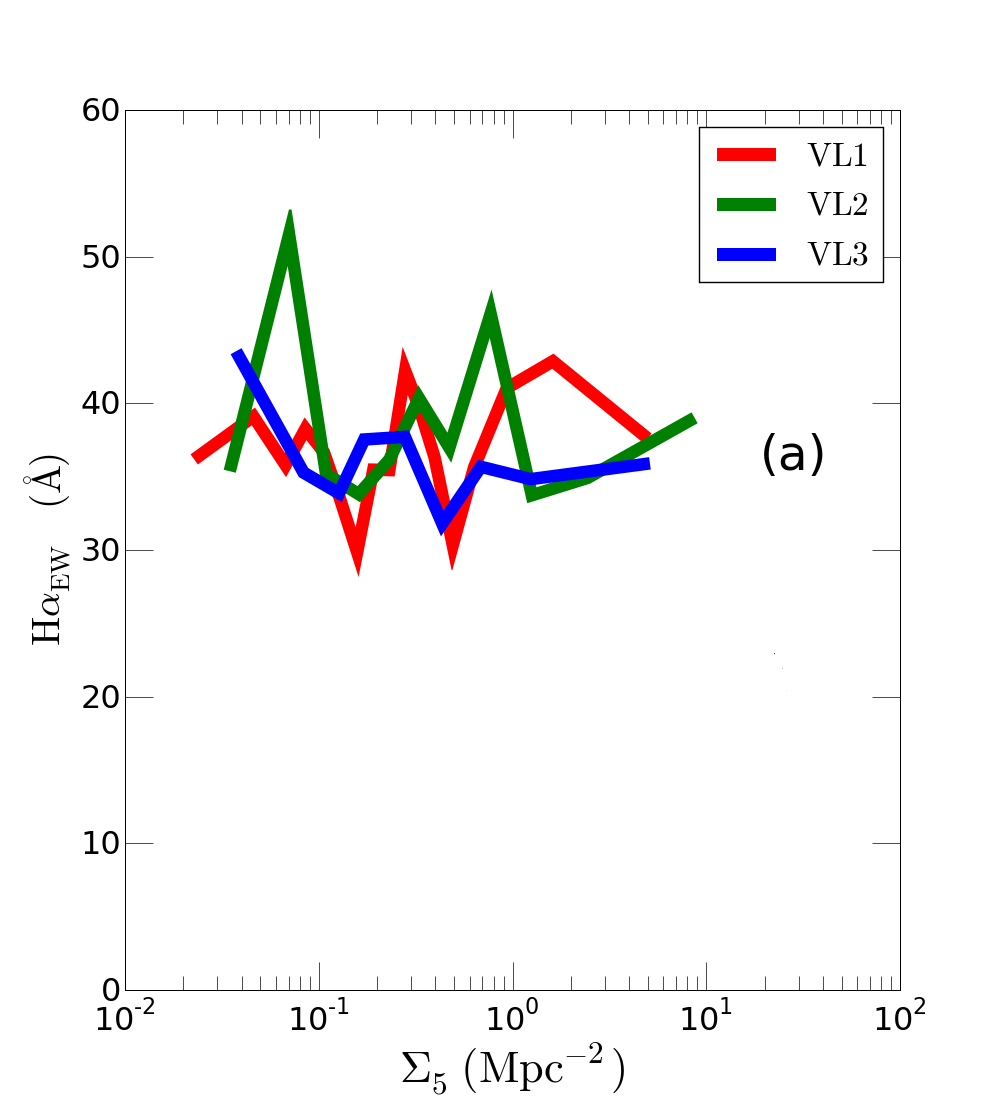}
\includegraphics[width=58mm, height=58mm]{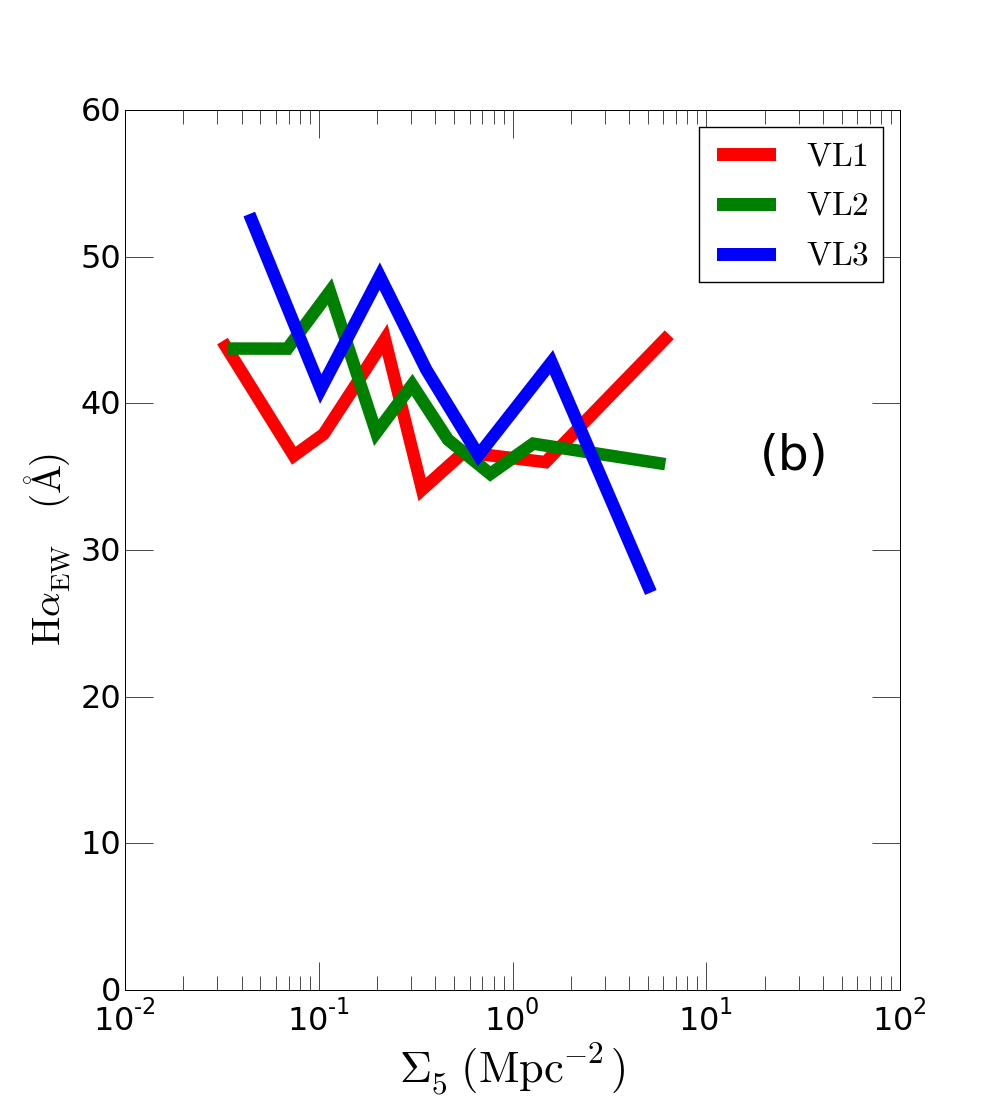}}
\caption{$\mathrm{EW}_{\mathrm{H}\alpha}$ as a function of density for the ``star-forming'' sample of 
galaxies for (a) early-types and (b) late-types in the three volume-limited samples (red: VL1, green: VL2, blue: VL3).
The lines represent the $75^{\mathrm{th}}$ percentile. Both panels use 50 galaxies per bin. Early star-forming galaxies do not show a trend between 
$\mathrm{EW}_{\mathrm{H}\alpha}$ and $\Sigma_{5}$ while late types show a weak correlation between the two parameters.}
\label{fig:sf}
\end{figure*}

The ``star-forming'' sample (Figure~\ref{fig:sf}), however, shows a weak or no relation between EW and density for either late or early-type star-forming galaxies, consistent with the results from the
previous sections. Both these results add further weight to the earlier conclusion that it is the passive 
galaxy population that causes the observed trend between EW or SFR and density. This highlights that it is not just the passive galaxies but the passive early-type galaxies that are largely 
responsible for the trends found by G\'{o}mez et al. (2003) and \citet{Lws:02}. 

\section{Discussion}

The comparison to the results of G\'{o}mez et al. (2003) when comparing EW with density is encouraging, with a remarkable similarity in the results from our sample for a
comparable $M_{r}$ range at a slightly higher redshift. This result also
gives confidence in our subsequent analysis.

\citet{Lws:02} and G\'{o}mez et al. (2003) proposed that suppression of SFR in dense environments is due to a physical process occurring 
in these environments that exceeds the reduction in SFR due to the changing population mix. Our results demonstrate, in contrast and consistent with the work of Weinmann et al. (2010), 
Bolzonella et al. (2010) and Wetzel et al. (2011),
that the reduction in the distribution of SFR in dense environments is primarily a
consequence of the increasing proportion of passive galaxies.
The absence of a trend between EW and density for the ``star-forming'' sample, visible in all three volume-limited samples, indicates that the galaxies that contribute to the
EW-density trend in the ``full'' sample are the passive population.



We see no evidence for changes in the distribution of SFRs (within the ``star-forming'' population) as a function of density.
There are no subsets of the population
where a SFR-density relation is apparent. Combined with Figure~\ref{fig:density_haew} this indicates that the absence of the SFR-density relationship is purely a 
consequence of selecting a star-forming population of galaxies
in agreement with Baldry et al. (2006) and McGee et al. (2011). 
A similar result was obtained by \citet{DD:06} who found that the SFR of late-type galaxies selected by neutral [H{\sc i}] gas content was not dependent on density.



Our final analysis focuses on the significance of galaxy morphologies on the SFR-density relationship. We show that early-type galaxies drive the EW-density and 
SFR-density relation to a far greater extent than late-type galaxies.
The early-types show an EW-density and SFR-density relationship that resembles the result for the ``full'' sample very closely, 
indicating that the passive early-types are indeed the driver of the relationship observed by G\'{o}mez et al. (2003).
This confirms that it is primarily the morphology-density relation that underpins the SFR-density relation, consistent with 
Baldry et al. (2006), Bolzonella et al. (2010), Thomas et al. (2010), Wetzel et al. (2011), Deng et al. (2011) and Lu et al. (2011) among others.



We can compare with the work of Wolf et al. (2009), Vulcani et al. (2010) and Li et al. (2011) through our analysis of the EW-density and SFR-density relation as a function of stellar
mass, as shown in Figures~\ref{ew_mass_bins} and \ref{sfr_mass_bins}. Von der Linden et al. (2010) and Li et al. (2011) observe a relation between SFR and density at low masses, Vulcani et al. (2010) 
observe such a relation across a range of masses and Wolf et al. (2009) observe the same relation at higher masses (but not for low mass galaxies). Figures~\ref{ew_mass_bins} and \ref{sfr_mass_bins} 
clearly show that we do not observe an SFR-density relation for star forming galaxies at any mass.

In Figure~\ref{sfr_hist} we show that for any mass the SFR distribution is the same for both high and low density. If there was a low SFR population of galaxies in high densities not observed 
in the field then we would expect to see this as a tail to lower SFRs in the higher density samples. We do not observe such a scenario for our sample of galaxies.

We stress that the lack of a SFR-density relation in our analysis does not rule out the presence of SFR suppression in star-forming galaxies. Balogh et al. (2004), using galaxy
colours, and McGee et al. (2011), using SSFRs, have also demonstrated a lack of an SFR-density relation. They both suggest that any suppression process must have
a very short timescale. McGee et al. (2011) also proposes that infall and quench may be a viable candidate for this rapid suppression and that observations of ``green valley'' transition galaxies
(McGee et al. 2009, Balogh et al. 2011) will aid in confirming the nature of the rapid SFR suppression processes.





There may also be some dependence on the environment metric used and there is work currently underway to investigate any effect arising through the use of different measures of environment
(Muldrew et al. 2011, Brough et al. in prep.) such as group masses \citep{Rob:11,Grts:11}.

We note that \citet{Pas:09} and \citet{vdB:08} have advanced similar arguments, relating to the distribution of SFR with density, based on slightly different analyses. In particular, 
these authors have argued in their analyses for the importance of
distinguishing between satellite and central galaxies. In this context, we note that we find no evidence for density-dependent variations in the joint SFR-mass distribution, 
even without distinguishing between satellites and centrals.  This raises the possibility that the satellite/central distinction is responsible for, or coincident with, the increasing fraction of 
passive and/or early-type galaxies for higher stellar masses and denser environments.  
There is also the possibility that we could be largely observing centrals, particularly at low surface densities, and if this is the case 
we may not expect to see environmental influences that would mainly act on satellite galaxies (e.g., ram pressure stripping).
Further detailed study of the relation between these results may be fruitful.

The question now becomes whether star-forming galaxies entering dense environments are quenched rapidly enough that we do not see any population in a low-SFR state (the ``infall-and-quench'' model),
or whether the morphology-density relation arises through density-dependent evolution, an ``in-situ evolution'' model, with similarities to the ``mass quenching'' model of Peng et al. (2010).
Such an ``in-situ'' model would be a modification to the
``downsizing by mass'' paradigm \citep{Cow:96} or the ``staged evolution'' paradigm \citep{Nsk:07}. In this variation, galaxies in dense environments would evolve faster than galaxies in
low-density environments, building their stellar mass faster and earlier, leading to the observed morphology-density relation, and consistent with the measured SFR-density relations at both low and
high redshift.

This ``in-situ'' evolution differs from the ``mass-quenching'' versus ``environmental quenching'' model of Peng et al. (2010), as galaxies of common mass would evolve differently in different
environments, in order to give rise to the observed population mix. In particular, the different star-forming fraction at a given mass as a function of density is a more natural outcome of the
``in-situ'' model model.

\section{Conclusion}
We have carried out an analysis of the impact of galaxy environments and stellar mass on SFR and EW for three volume-limited samples from the GAMA survey out to $z=0.2$.
The galaxies populate both field and cluster environments, and encompass star-forming and passive, early and late galaxy types. 
We investigate the relation between SFR as well as $\mathrm{EW}_{\mathrm{H}\alpha}$ and environment, initially to confirm the trends shown by G\'{o}mez et al. (2003). Subsequently we 
incorporate stellar mass to understand the lack of any trend between SFR or $\mathrm{EW}_{\mathrm{H}\alpha}$ and density for the star-forming population.
The absence of a trend implies the possibility of two evolutionary trends where either the galaxies undergo ``in-situ'' evolution (Baldry et al. 2004, Balogh et al. 2004)
or the galaxies do ``infall'' and are subsequently quenched within a short period of time, particularly for low-mass galaxies 
(Bolzonella et al. 2010, Wetzel et al. 2011, Weinmann et al. 2010).

We show that there is no strong effect due to density on galaxy SFRs for the star-forming population.
Furthermore, we show that SFR is largely dependent on stellar mass rather than density, for a variety of samples binned according to stellar mass, density, SFR and EW.
The investigation into morphology clearly shows that it is the increasing fraction of passive early-type galaxies that are the largest contributor to the suppression in SFR at high densities.
In contrast, the distribution of SFRs for star-forming galaxies shows no change with density, and no evidence for a physical suppression in dense environments. 
The combination of these facts forces us to conclude that the SFR-density relation that we observe for galaxies as a whole 
is largely due to the proportion of passive early-type galaxies present in the clusters,
as opposed to some physical process acting on star-forming galaxies to suppress their SFRs as they enter a cluster. Such processes clearly occur, and have been observed to affect individual
systems in nearby clusters, but they may not be a dominant effect in governing galaxy evolution. 

A similar result was found by \citet{Bal:04} where they identified that while there is strong and continuous variation between the relative numbers for quiescent and star-forming 
galaxies there is little correlation between the distribution of SFRs and density for star-forming galaxies. \citet{Bal:04} also uses the $5^{\mathrm{th}}$ nearest neighbour (as well as a 
three-dimensional density estimator) as a measure of density and the $\mathrm{EW}_{\mathrm{H}\alpha}$ as a measure of star-forming activity. Recent work by Peng et al. (2010) also shows that
the SFR is primarily dependent on the mass of galaxies and any evolutionary effects are secondary.


There are several key issues that remain unresolved. This includes in particular whether the timescales on which star-formation can be suppressed by the commonly proposed mechanisms of in-fall and
quench, harassment, or ram-pressure stripping, are short enough that they remain a viable explanation to support an ``infall-and-quench'' model for the morphology-density relation.
\citet{Bal:04,Bam:08} and \citet{NB:11} have argued that this is a viable explanation for the observed suppression of SFR.
An alternative scenario, which deserves further investigation is a model of ``in-situ'' evolution where the passive early-types may have evolved early and rapidly within the dense environments, 
while star-forming galaxies evolve more slowly in the field. This model is not dissimilar to the ``staged evolution'' model of \citet{Nsk:07}, although it invokes a density-dependence on
the timescale of the evolution in addition to the mass-dependence. We plan to explore such a model in more detail in future work.

\section*{Acknowledgments}
The authors would like to thank the referee for the insightful comments that were provided, putting the paper in a broader and a more relevant context.
D.B.W acknowledges the support provided by the University of Sydney Postgraduate Award.
PN acknowledges a Royal Society URF and an ERC StG grant (DEGAS-259586).
GAMA is a joint European-Australasian project based around a spectroscopic campaign using the Anglo-Australian Telescope. 
The GAMA input catalogue is based on data taken from the Sloan Digital Sky Survey and the UKIRT Infrared Deep Sky Survey. 
Complementary imaging of the GAMA regions is being obtained by a number of independent survey programs including GALEX MIS, 
VST KIDS, VISTA VIKING, WISE, Herschel-ATLAS, GMRT and ASKAP providing UV to radio coverage. GAMA is funded by the STFC (UK), 
the ARC (Australia), the AAO, and the participating institutions. The GAMA website is http://www.gama-survey.org/ .

\label{lastpage}

\end{document}